\documentclass[12pt,preprint]{aastex}

\shorttitle{CAUSAL BACKREACTION w/RECURSIVE NONLINEARITY}
\shortauthors{BOCHNER}

\begin{document}

\title{COSMIC ACCELERATION FROM CAUSAL BACKREACTION 
	     WITH RECURSIVE NONLINEARITIES}

\author{Brett Bochner}
\affil{Department of Physics and Astronomy, 
Hofstra University, Hempstead, NY 11549}
\email{brett\_bochner@alum.mit.edu, phybdb@hofstra.edu}

\begin{abstract}
We revisit the causal backreaction paradigm, in which the 
need for Dark Energy is eliminated via the generation of 
an apparent cosmic acceleration from the causal flow of 
inhomogeneity information coming in towards each observer 
from distant structure-forming regions. A second-generation 
version of this formalism is developed, now incorporating 
the effects of ``recursive nonlinearities'': the process 
by which metric perturbations already established by some 
given time will subsequently act to slow down all future 
flows of inhomogeneity information. In this new formulation, 
the long-range effects of causal backreaction are damped, 
substantially weakening its impact for simulated models 
that were previously best-fit cosmologies. Despite this 
result, we find that causal backreaction can be recovered 
as a replacement for Dark Energy through the adoption 
of larger values for the dimensionless `strength' of the 
clustering evolution functions being modeled -- a change 
justified by the hierarchical nature of clustering and 
virialization in the universe, occurring as it does on 
multiple cosmic length scales simultaneously. With this 
and with the addition of one extra model parameter used to 
represent the slowdown of clustering due to astrophysical 
feedback processes, an alternative cosmic concordance can 
once again be achieved for a matter-only universe in which 
the apparent acceleration is generated entirely by causal 
backreaction effects. The only significant drawback is a 
new degeneracy which broadens our predicted range for the 
observed jerk parameter $j_{0}^{\mathrm{Obs}}$, thus 
removing what had appeared to be a clear signature for 
distinguishing causal backreaction from Cosmological 
Constant $\Lambda$CDM. Considering the long-term fate 
of the universe, we find that incorporating recursive 
nonlinearities appears to make the possibility of an 
`eternal' acceleration due to causal backreaction 
far less likely; though this conclusion does not take 
into account potential influences due to gravitational 
nonlinearities or the large-scale breakdown of 
cosmological isotropy, effects not easily modeled 
within this formalism. 
\end{abstract}


\section{\label{SecIntro}INTRODUCTION: COSMIC CONCORDANCE 
                         AND CAUSAL BACKREACTION} 

One of the key questions in Cosmology today relates to the 
still-unsolved problem of what is causing the observed cosmic 
acceleration. Primarily indicated by Hubble curves constructed 
from luminosity distance measurements of Type Ia supernovae 
\citep{PerlAccel99,RiessAccel98}, this (possibly apparent) 
acceleration is just one aspect of the struggle for a consistent 
picture of the universe; a picture that would also require an 
explanation of the gap between the observed clustering matter 
content of $\Omega _\mathrm{M} \sim 0.3$ 
\citep[e.g.,][]{TurnerCaseOm0pt33} and the value of 
$\Omega _\mathrm{Tot} = 1$ as indicated by Cosmic Microwave 
Background measurements of spatial flatness 
\citep{WMAP7yrCosmInterp}, as well as a solution of the 
``Age Problem/Crisis'' for matter-only (i.e., decelerating) 
cosmologies in which the universe appears to be younger than 
some of its oldest constituents \citep[e.g.,][]{TurnerCosmoSense}, 
along with explanations of other important issues. 

The standard approach to solving these problems is to introduce 
some form of ``Dark Energy'' -- the simplest case being the 
Cosmological Constant, $\Lambda$ -- which can fill the gap via 
$\Omega_\mathrm{DE} = \Omega_\mathrm{Tot} - \Omega_\mathrm{M} \sim 0.7$, 
which possesses negative pressure in order to achieve cosmic acceleration 
\citep[e.g.,][]{KolbTurner}, and which (for non-$\Lambda$ cases) 
recruits some form of internal nonadiabatic pressure 
\citep[e.g.,][]{CaldDDEsNotSmooth} in order to avoid clustering 
as matter does. Thus the introduction of Dark Energy (often using 
spatially-flat $\Lambda$CDM models) has led to a broadly-consistent 
``Cosmic Concordance'' -- an empirical outline which has seemed so 
far to succeed fairly well \citep[e.g.,][]{WMAP7yrCosmInterp} at 
developing into a consistent cosmological picture. 

There are serious aesthetic problems with Dark Energy, however, as is 
well known; the most obvious being the problematical introduction of 
a completely unknown substance as the dominant component of the universe. 
Beyond that, a static (i.e., Cosmological Constant) form of Dark Energy 
suffers from two different fine-tuning problems: one being the issue that 
$\rho _{\Lambda}$ is $\sim$$120$ orders of magnitude smaller than what 
would be expected from the Planck scale \citep{KolbTurner}; and the other 
being a ``Coincidence Problem" \citep[e.g.,][]{ArkaniHamedCoinc}, 
questioning why observers {\it today} happen to live so near the onset 
of $\Lambda$-domination, given 
$\rho _{\Lambda}/ \rho _{\mathrm{M}} \propto a^{3}$. Moving to a 
dynamically-evolving Dark Energy (DDE), however, invites other problems, 
since the {\it self-attractive} nature of negative-pressure substances 
(i.e., $\partial E / \partial V = - P > 0$) means that a DDE may cluster 
spatially \citep{CaldDDEsNotSmooth}, possibly ruining it as a 
``smoothly-distributed'' cosmic ingredient. This could potentially be 
solved through the ad-hoc addition of some form of nonadiabatic support 
pressure for the DDE \citep{HuGDM98}, but this is a possibility which 
we have argued against elsewhere \citep{BochnerAccelPaperI} on 
thermodynamically-based cosmological grounds. 

Besides Dark Energy, various other methods have been used to attempt to 
explain the observed acceleration, such as employing modified gravity 
\citep[e.g.,][]{TroddenModGravAccel}, or assuming the existence of an 
underdense void centered not too far from our cosmic location 
\citep[e.g.,][]{TomitaSNeVoid}. But to avoid the substantial (and perhaps 
needless) complications which arise when assuming departures from General 
Relativity, as well as the non-Copernican `specialness' implied by a local 
void, we will instead use the feedback from cosmological structure formation 
itself as a {\it natural} trigger for the onset of acceleration -- a trigger 
that automatically activates at just the right time for observers to see 
it, due to the fact that all such observers will have been created by that 
very same structure formation which generates the observed cosmic acceleration. 

This approach, known generally as ``backreaction'', was used by this author 
in \citet{BochnerAccelPaperI} (henceforth BBI) to find several clustering 
models which managed to precisely reproduce the apparent acceleration seen 
in Hubble curves of Type Ia supernova standard candles, while simultaneously 
driving a number of important cosmological parameters to within a close 
proximity of their Concordance values -- including the age of the universe, 
the matter density required for spatial flatness, the present-day deceleration 
parameter, and the angular scale of the Cosmic Microwave Background.

The ability of our models to achieve these goals, despite the generally 
pessimistic view of backreaction typically held by researchers currently 
\citep[e.g.,][]{SchwarzBackReactNotYet}, was due to our adoption of an 
explicitly causal variety of backreaction, which admits the possibility 
of substantial backreaction from Newtonian-strength perturbations. The 
standard formalism used for computing backreaction effects, developed 
through the extensive work of Buchert and collaborators 
\citep[e.g.,][]{BuchertEhlers97,BuchertKerschSicka}, is non-causal in 
the sense that it drops all `gravitomagnetic' (i.e., velocity-dependent) 
effects, thus rendering it unable to account for metric perturbation 
information flowing (at the speed of null rays) from structures forming 
in one part of the universe, to observers in another. In similar fashion, 
typical studies of cosmic structure formation are also non-causal in that 
they use the Poisson equation without time derivatives of the perturbation 
potential, thus computing metric perturbations from {\it local} matter 
inhomogeneities only, disregarding all gravitational information coming 
in from elsewhere in space. The result is a mistaken (but widespread) 
notion that the entire Newtonian backreaction $Q_{N}$ can be expressed 
mathematically as a total divergence, thus ultimately rendering it 
negligible. But by restoring causality with a ``causal updating'' 
integral that incorporates perturbations to an observer's metric coming 
from inhomogeneities all the way out to the edge of their observational 
horizon, we find (BBI) that the sum of such `innumerable' 
Newtonian-strength perturbations -- which increase in number as $r^{2}$ 
within a spherical shell at distance $r$ from the observer, more than 
compensating for their $1/r$ weakening with distance -- adds up to a 
total backreaction effect that is not only non-negligible (regardless 
of the smallness of $v^{2}/c^{2}$ for most matter flows), but is in fact 
a dominant cosmological effect that is fully capable of reproducing 
the observed cosmic acceleration in a fully `concordant' manner. 

Despite these successes, a major problem with our model in BBI is its 
utter simplicity: it is clearly a toy model, with the results presented 
there serving primarily as `proof-of-principle' tests, rather than as 
precision cosmological predictions. Though the simplifications of the 
model are many, one in particular is serious in its consequences, while 
fortunately being not too difficult to fix: specifically, this is the 
dropping of what we have termed ``recursive nonlinearities''. Unrelated 
to {\it gravitational} nonlinearities, or to the nonlinear regime of 
density perturbations in structure formation, recursive nonlinearities 
embody the fact that the integrated propagation time of a null ray 
carrying perturbation information to an observer from a distant 
virializing structure would itself be affected by all of the other 
perturbation information that has already come in to cross that ray's 
path from everywhere else, during all times prior to arrival. In other 
words, causal updating is itself slowed by the metric perturbation 
information carried by causal updating, creating an operationally 
nonlinear problem. 

This issue was necessarily neglected in BBI, as that work was devoted to 
introducing our `zeroth-order' approach to causal backreaction. But here 
we fix this problem, incorporating recursive nonlinearities into a new, 
`first-order' version of our phenomenological model. We will find that 
this alteration significantly changes our results, causing a profound 
weakening of the backreaction effects generated by a given level of 
clustering, as well as significantly damping the long-term effects of 
information from old perturbations coming in from extreme distances. In 
order to retain causal backreaction as a viable model for generating the 
observed cosmic acceleration -- presuming here that this should indeed 
be done -- it will be necessary to re-interpret the meaning of our 
(inherently empirical) `clumping evolution functions' to now consider 
the effects of hierarchical clustering on a variety of cosmic scales. 
Doing this, we will show that a successful alternative concordance 
can once again be achieved, with the right amount (and temporal behavior) 
of acceleration, and with good cosmological parameters. 

This paper will be organized as follows: in Section~\ref{SecRecNonlinForm}, 
we will re-introduce our original causal backreaction formalism, and 
then describe the changes implemented in order to incorporate recursive 
nonlinearities into the model. In Section~\ref{SecRecNonlinResults}, 
we will explore the results of the new formalism, and discuss the 
implications of the model parameters that are now needed to achieve 
good data fits. Furthermore, we will discuss how the damping effects 
due to recursive nonlinearities would alter the key factors that 
determine the `ultimate' fate of the universe, as was discussed 
for our original formalism in BBI, given an acceleration driven by 
causal backreaction rather than by some form of Dark Energy. 
Finally, in Section~\ref{SecSummConclude}, we conclude with a summary 
of these ideas and results, highlighting the role of causal backreaction 
as a fundamental component of cosmological analysis and modeling.

\section{\label{SecRecNonlinForm}THE CAUSAL BACKREACTION FORMALISM: 
                            OLD METHODS AND NEW DEVELOPMENTS} 

\subsection{\label{SubSecOldApprox}The Original Toy Model, and its 
Approximations and Simplifications} 

We recall here that the basic premise of the formalism developed in BBI is to 
phenomenologically represent the physical processes of structure formation 
-- complex even at the level of Newtonian-strength gravitational perturbations 
-- in a simple and convenient way. The physics at work within most clustering 
masses should be as follows: collapsing overdensities stabilize themselves and 
halt their collapse by concentrating their local vorticity (or equivalently, 
by creating a large local velocity dispersion); this concentrated vorticity or 
velocity dispersion leads to real, extra volume expansion in accordance with 
the Raychaudhuri equation \citep{HawkingEllis}, representable (in the final 
state) at great distances by the tail of a Newtonian potential perturbation 
to the background Friedman Robertson-Walker (FRW) metric; and this Newtonian 
tail propagates causally outward into space by inducing inward mass flows 
towards the virialized object from farther and farther distances as time 
passes. The total perturbation at time $t$ for any location in space -- which 
will be independent of position, assuming similar structure formation rates 
everywhere -- will then be the combined effects of innumerable 
Newtonian tails of this type, coming in towards the observer from the 
virializing masses (in all directions) which by that observation time 
have entered within the observer's cosmological ``clustering horizon''. 

As is well known, the expansion evolution (i.e., the Friedmann equation) 
for some spherical volume $\mathbf{V}$ can be derived -- using 
nonrelativistic Newtonian equations, in fact, for a matter-dominated universe 
\citep[][pp. 474-475]{WeinbergGravCosmo} -- without reference to anything 
outside of that sphere. In contrast, the Newtonian-level backreaction terms 
which we utilize here are due to perturbation information coming in from 
structures located predominantly outside of $\mathbf{V}$ (since one must go 
to cosmological distances for the effects to add up significantly). For any 
condensed structure (at great distance) which provides a gravitational pull 
upon the mass in $\mathbf{V}$, its main perturbative effect is simply to impose 
an extra (Newtonian) perturbation potential upon $\mathbf{V}$ as an addition 
to its original cosmological metric. Our phenomenological approach, therefore, 
is one in which we model the inhomogeneity-perturbed evolution of $\mathbf{V}$ 
with a metric that contains the individually-Newtonian contributions to 
the perturbation potential within $\mathbf{V}$ (``$\Phi _\mathbf{V} (t)$'') 
from all clumped, virialized structures outside of $\mathbf{V}$ that have 
been causally `seen' within $\mathbf{V}$ by time $t$, superposed 
{\it on top of} the background Friedmann expansion of $\mathbf{V}$.

We will reiterate the mathematical essentials of this formalism below, 
in Section~\ref{SubSecOldFormalism}; but first we must recount the 
various approximations and simplifications which have gone into our 
analysis, to consider their importance and the feasibility of 
eliminating them in order to develop a greater degree of 
physical realism in these causal backreaction models. 

First of all, though our backreaction-induced metric perturbations 
will indeed be time-dependent (due to the causal flow of inhomogeneity 
information), they will be entirely spatially-{\it independent}. 
As noted above, we do not seek to achieve an observed acceleration 
through the mechanism of a local void; but going even further, our 
model does not explicitly include any spatial variations whatsoever. 
Rather, the system being modeled is what we term a 
``smoothly-inhomogeneous'' universe, in which all perturbation 
information blends together evenly in a way that is essentially 
independent of cosmic position. 

Now, this simplification is one made out of practical necessity, not 
physical realism. The smoothly-inhomogeneous approximation relies 
upon an assumption of randomly-distributed clustering -- which is 
certainly not true, as large clusters are not independent of each other, 
but preferentially clump near one another and are mutually correlated 
-- and this becomes ever less true during the ongoing cosmic evolution, 
as the universe grows more inhomogeneous with time. Furthermore, this 
simplification relies upon the assumption that the region of space 
responsible for the dominant contributions to causal backreaction 
within volume $\mathbf{V}$ will be large enough to contain a 
cosmologically-representative sample of both clusters and voids; 
but as we will see below in Section~\ref{SecRecNonlinResults}, 
adding in recursive nonlinearities (to correct another simplification, 
as described below) greatly reduces the size of the cosmological region 
affecting $\mathbf{V}$ from what it was in our original toy model, 
potentially calling this assumption into question. 

A proper accounting of causal backreaction in a realistically 
inhomogeneous universe would require the implementation of a fully 
spatially-detailed, three-dimensional cosmic structure simulation 
program -- perhaps along the lines of \citet[]{VIRGOsims}, for example 
-- but with Newtonian-level backreaction effects from causal updating 
now added in. The development of such a 3D simulation model is far 
beyond the scope of this paper (and beyond the efforts of any 
individual researcher), but would be a useful mission to be 
undertaken by the cosmological community at large. 

A second simplification is our use of Newtonian potential terms -- 
i.e., the long-distance approximation of the Schwarzschild metric 
\citep[e.g.,][]{WeinbergGravCosmo} -- to represent 
the tail of each individual perturbation felt from far away, rather than 
using a long-distance approximation of the Kerr metric for spinning masses 
\citep{RefKerrBH}, despite the crucial role of some form of vorticity in 
stabilizing most structures against singular collapse. In this case, however, 
the approximation is a good one. A relatively small amount of vorticity can 
suffice for the self-stabilization of a clumped mass, if it is applied 
perpetually; and the specific angular momentum (i.e., $[J/(Mc)]$) will be 
small for any mass not on the verge of being an extremal black hole. 
Furthermore, it is easy to show \citep[e.g.,][]{FranklinKerrPert} that the 
highest-order deviations from the Newtonian expression in the diagonal metric 
components will go like $[J/(Mcr)]^{2}$, thus being entirely negligible at 
the huge distances relevant for causal backreaction. (And the leading-order 
{\it off-diagonal} Kerr perturbation terms, though actually proportional 
to $(J/r)$, will effectively cancel out due to angle-averaging in the 
smoothly-inhomogeneous approximation, as discussed in BBI.) Thus it appears 
quite safe to ignore any Kerr-specific perturbation effects for physically 
reasonable situations. 

Third, our formalism neglects the purely observational effects of localized 
inhomogeneities, such as lensing along beam paths \citep[e.g.,][]{KantowSwiss03} 
for rays from standard candles, and similar perturbative effects upon the 
apparent luminosity distance relationship for rays passing through inhomogeneous 
regions, which in some models represents the primary `backreaction' effect 
resulting from structure formation \citep[e.g.,][]{BisManNotAppAccel}. 
Even if such effects by themselves are too small to generate an observed cosmic 
acceleration, they will still alter the output parameters estimated while 
using any cosmological model (including ours), and thus should be kept track of; 
and in case our causal backreaction method also falls short of providing the 
full result of an apparent acceleration all by itself, it might successfully 
be combined with these other observational effects upon the light rays 
to produce an apparent acceleration once everything is added together 
(this point to be discussed again in Section~\ref{SubUnityClumpWeak}). 
Combining these purely observational effects with those from 
causal backreaction is therefore an important task, and likely quite a 
feasible one; though not one addressed yet in this current paper. 

The next, most theoretically treacherous approximation is our neglect 
of nonlinear gravitational effects, a simplification made implicitly by  
our method of linearly adding together the individual metric perturbations 
contributed by different self-stabilized mass `clumps' in order to produce 
the total, summed, Newtonian-strength perturbation potential. (Note that this 
is not the full ``Newtonian'' approximation usually employed, since while we 
do assume weak-gravity, we do not completely assume `slow-motion' -- in the 
sense of dropping all time derivatives of the perturbation potential -- as that 
would neglect the causal flow of perturbation information.) Thus our formalism 
explicitly neglects the nonlinear, purely general-relativistic effects that 
most other researchers primarily focus upon when studying ``backreaction''. 
This approximation becomes increasingly bad as the magnitude of the summed 
perturbation potential approaches unity; but since (as will be seen below) 
this potential typically does not grow to values in excess of $\sim$$0.5-0.6$ 
or so as $t \rightarrow t_{0}$ for most of our best-fitting simulation runs, 
the approximation is probably good enough for our simulations to provide fairly 
accurate estimations of the cosmic evolution up to now, and of our measurable 
cosmological parameters. (And to the extent that it is not good enough, a 
significant contribution due to nonlinear gravitational terms would likely 
only help produce the desired acceleration even more easily.) Thus it is 
probably not necessary for us to include higher-order gravitational terms 
in our formalism, in order to achieve a sufficiently reliable understanding 
of the currently-observable universe for our present purpose of pointing the 
way towards an alternative concordance; a fortunate situation, since our model 
is fundamentally designed around a linearized-gravity approach, and it may be 
challenging to find any convenient way of modifying it to include nonlinear 
gravitational effects. On the other hand, given the ever-increasing strength 
of gravitational nonlinearities in the cosmos over time, a fully 
general-relativistic model of causal backreaction (computed using a 
3D simulation of realistically-distributed inhomogeneities) would almost 
certainly be necessary for accurately predicting the long-term future 
evolution of the universe. 

Lastly, there is the approximation regarding what we have referred to 
as ``recursive nonlinearities''. As will be seen from the metric given 
below in Section~\ref{SubSecOldFormalism}, one of the effects of causal 
backreaction is real extra volume creation. But since causal backreaction 
depends upon the propagation of inhomogeneity information through space, 
the extra volume produced by old information from perturbations will 
slow down the propagation of all future inhomogeneity information 
(as well as carrying all perturbing masses farther away from all 
observation points), thus feeding back upon the causal backreaction 
process in such a way as to strongly dampen it. Of all of the 
simplifications and approximations discussed so far in this subsection, 
the neglect of these recursive nonlinearities most likely has the 
strongest impact upon the quantitative predictions emerging from our 
causal backreaction models. Fortunately, however, fixing this problem 
by adding these recursive nonlinearities into our formalism is one of 
the simpler improvements in physical realism for us to make; and hence, 
this paper focuses upon achieving this fix, and then calculating and 
interpreting the results produced by this `second-generation' 
causal backreaction formalism.

\subsection{\label{SubSecOldFormalism}The Old Formalism and its Results} 

Here we recall the technical details of our original formalism 
developed in BBI, to set the stage for its further development to follow. 

To obtain the Newtonian approximation of a single `clumped' 
(i.e., virialized, self-stabilized) object of mass $M$, embedded at the 
origin ($r = 0$) in an expanding, spatially-flat, matter-dominated (MD) 
universe, one may linearize the McVittie solution \citep{McVittieBHinFRW}, 
as can be seen from the perturbed FRW expression given in 
\citet{KaloKlebanBHinFRW}. The resulting Newtonianly-perturbed 
FRW cosmology is given by the metric: 
\begin{equation}
ds^2 \approx - c^{2} [1 + (2/c^{2}) \Phi (t)] d t^{2} 
+ [a_{\mathrm{MD}}(t)]^{2} [1 - (2/c^{2}) \Phi (t)] d r^{2}
+ [a_{\mathrm{MD}}(t)]^{2} r^{2} [d {\theta}^{2} 
                           + \sin^{2}{\theta} d {\phi}^{2}] 
~ , 
\label{NewtPertSingleClump}
\end{equation} 
where $\Phi (t) \equiv \{ - G M / [a_{\mathrm{MD}}(t) r] \}$, 
and $a_{\mathrm{MD}}(t) \propto t^{2/3}$ is the 
unperturbed MD scale factor evolution function. 

For our model of a smoothly-inhomogeneous universe, we assume a random 
distribution of clustered masses, being the same essentially everywhere 
and in every direction. So turning the above expression around, 
we consider the situation for an observer at the origin, whose metric is 
affected by a collection of (roughly identical) discrete masses -- for 
now confined to a spherical shell at coordinate distance $r^{\prime}$, 
and with {\it total} mass $M$ -- that are distributed randomly in direction. 
The different directions of the various clumps does not matter for the 
observer's $g_{t t}$ metric component; but it does matter for the spatial 
metric components, due to the fact that a spatial displacement from the 
origin would pick up (direction-dependent) factors of 
$\cos^{2} \theta$ in $ds^2$ from the different angles of the motion with 
respect to the individual clumped masses, since only the radial projection 
of a given translation (with respect to a particular clump) will `feel' the 
perturbation potential from that clump in its contribution (all within 
$g_{r r}$) to the interval $ds ^{2}$. Averaging over direction in three 
dimensions, the spatial metric terms therefore pick up a factor of 
$\langle \cos^{2} \theta \rangle = (1/3)$ for the total effect when summing 
over all of the discrete masses; and the total, (gravitationally-)linearly 
summed and angle-averaged metric for this observer at the origin can then 
be written in `isotropized' fashion, as: 
\begin{equation}
ds^{2} = 
- c^{2} \{ 1 - [ R_{\mathrm{Sch}}(t) / r^{\prime} ] \} ~ dt^{2} 
~ + ~ [a_{\mathrm{MD}}(t)]^{2} 
\{ 1 + (1/3) [ R_{\mathrm{Sch}}(t) / r^{\prime} ] \} 
~ \vert d \vec{r} \vert ^{2} ~ ,
\label{EqnAngAvgBHMatDomWeakFlat}
\end{equation} 
where 
$R_{\mathrm{Sch}}(t) \equiv \{ (2 G M / c^{2} ) / 
[a_{\mathrm{MD}}(t)] \}$, and 
$\vert d \vec{r} \vert ^{2} 
\equiv (d r^{2} + r^{2} d \theta ^{2} 
+ r^{2} \sin^{2}{\theta} d \phi ^{2} ) 
= \vert d \vec{x} \vert ^{2} 
\equiv (dx^{2} + dy^{2} + dz^{2})$. 
(Note that this factor of $1/3$ is not `fundamental', but is merely 
the result of our approximating the linearized sum of many individual 
`Newtonian' solutions, which effectively spreads out the total spatial 
perturbation among all three spatial metric terms, rather than confining 
it solely to $g_{r r}$, as is usual. Thus the general relativistic 
expectation of equal temporal and spatial potentials -- i.e., 
$\psi = \phi$ -- is not really violated here, and no actual new physics 
or modified gravity is implied by it.) We regard the term multiplying 
$[a_{\mathrm{MD}}(t)]^{2} \vert d \vec{r} \vert ^{2}$ as a `true' 
increase in spatial volume, and the $g_{t t}$ term as an `observational' 
term, slowing down the perceptions of observers at any given $t$. Thus, 
even if the spatial term by itself is not sufficient to generate a real 
volumetric acceleration, the spatial and temporal perturbations 
coupled together may indeed be enough to create an ``apparent acceleration" 
capable of explaining all relevant cosmological observations. 

The expression in Equation~\ref{EqnAngAvgBHMatDomWeakFlat} represents 
the perturbations to the metric due to masses at some specific coordinate 
distance $r^{\prime}$ -- and thus from a specific look-back time 
$t^{\prime}$ -- as seen from some particular observational point. The 
total metric at that spacetime point must be computed via an integration 
over all possible distances, out to the distance (and thus look-back time) 
at which the universe had been essentially unclustered. Finally, a light 
ray reaching us from its source (e.g., a Type Ia supernova being used as a 
standard candle) travels to us in a path composed of a collection of such 
points, where the metric at each point must be calculated via its own 
integration out to its individual ``inhomogeneity horizon''; and only by 
calculating the metric at every point in the pathway from the supernova 
to our final location here at $r = z = 0$ can we figure out the total 
distance that the light ray has been able to travel through the 
increasingly perturbed metric, given its emission at some specific 
redshift $z$. 

Finite look-back times imply that when one is feeling the effects of 
clumps at cosmological distances, what one is really sampling is the 
clustering as it was at an earlier, retarded time. Thus the cosmological 
evolution at all times is dependent upon the entire history of the 
development of clustering. In BBI, we defined a heuristic ``clumping 
evolution function'', $\Psi (t)$, intended to simply represent 
the fraction of the cosmic matter that has completed its relaxation 
to a self-stabilized, steady state by cosmic time $t$, as opposed to 
that mass still freely expanding (or collapsing) within a 
still dynamically-evolving patch of space. The function $\Psi (t)$ 
was therefore defined over the range from $\Psi (t) = 0$ (perfectly 
smooth matter), to $\Psi (t) = 1$ (everything clumped). This is an 
extremely simplified way of representing the virialization of 
cosmic structures, and we will in fact be forced to re-evaluate the 
meaning (and the range) of the function $\Psi (t)$ later on in this 
paper. But for now, we just recall that the specific functions used 
for $\Psi (t)$ in our numerical simulations are not rigorously 
derived from first principles, but rather are adopted for simplicity, 
with their functional forms and input parameters being motivated by 
physically reasonable structure formation behaviors, in conjunction 
with certain quantitative cosmological measurements. 

\begin{figure}
\begin{center}
\includegraphics[scale=0.75]{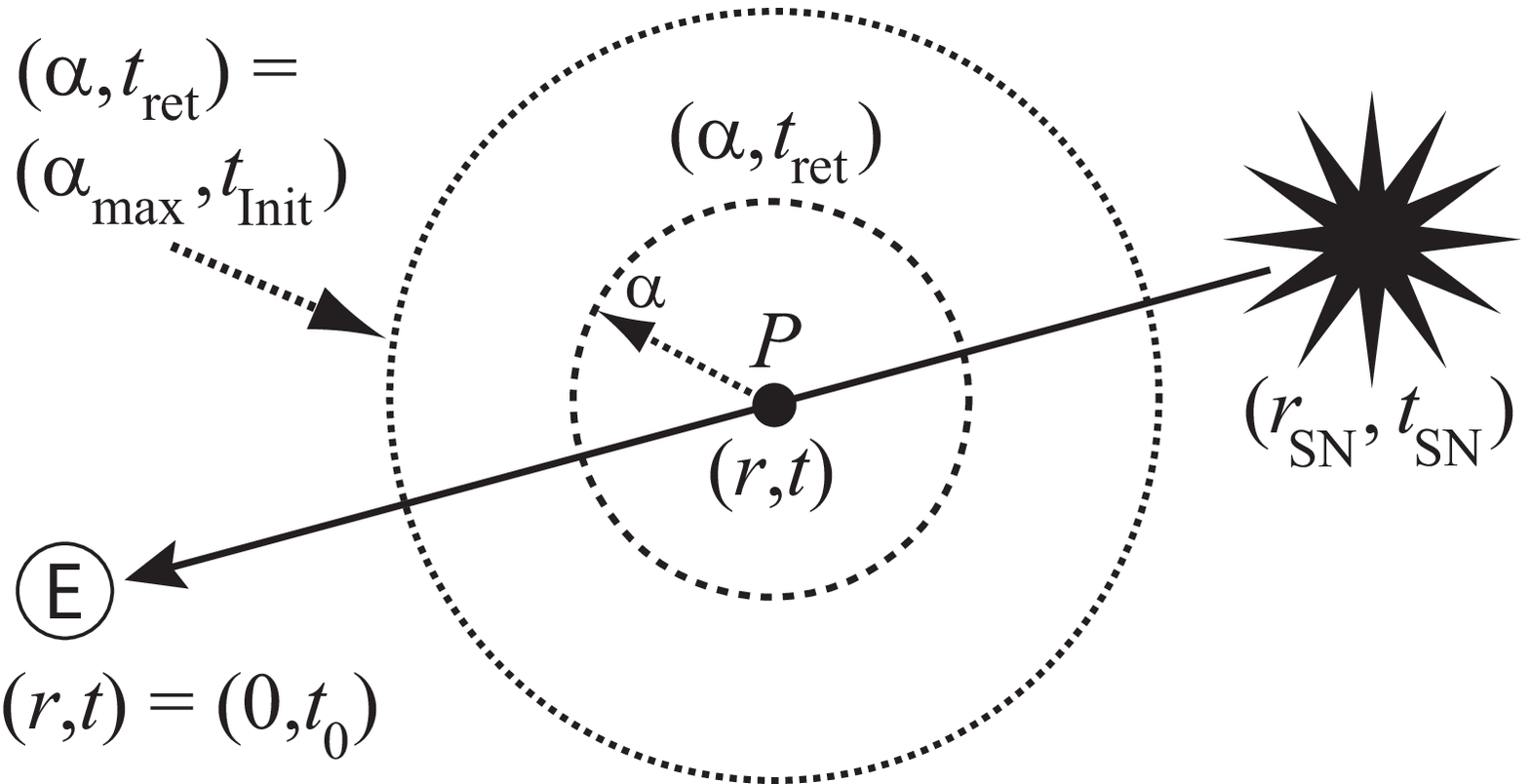}
\end{center}
\caption{Geometry for computing the inhomogeneity-perturbed 
metric at each point along the integrated path of a light ray 
from a supernova to our observation point at Earth.}
\label{FigSNRayTraceInts}
\end{figure}

Now, to determine how Hubble curves obtained from 
standard candle measurements are calculated in our formalism, 
consider a light ray emitted by a supernova at cosmic 
coordinate time $t = t_{\mathrm{SN}}$, which then propagates 
from the supernova at $r = r_{\mathrm{SN}}$, to us at $r = 0$, 
$t = t_{0}$. We refer here to the geometry depicted in 
Figure~\ref{FigSNRayTraceInts}.

\placefigure{FigSNRayTraceInts}

For each point $P \equiv (r,t)$ of the trajectory, 
the metric at that point will be perturbed away from the 
background FRW form by all of the virialized clumps that have 
entered its causal horizon. Consider a sphere of (coordinate) 
radius $\alpha$, centered around point $P$, with coordinates 
$(\alpha, t_{\mathrm{ret}})$ (where $t_{\mathrm{ret}} \leq t$ 
is the retarded time), defined such that the information 
about the state of the clumping of matter on that sphere 
at time $t_{\mathrm{ret}}$ will arrive -- via causal 
updating, traveling at the speed of null rays -- to 
point $P$ at the precise time $t$. To compute the 
fully-perturbed metric at $P$, we must integrate over the 
clumping effects of all such radii $\alpha$, from 
$\alpha = 0$ out to $\alpha _{\mathrm{max}}$, the farthest 
distance from $P$ out from which clumping information can 
have causally arrived since the clustering of matter had 
originally begun in cosmic history. 

For the remaining calculations in this subsection, note that 
we will be using our old formulation from BBI, {\it without} 
recursive nonlinearities (those changes will be presented later, 
in Subsection~\ref{SubSecNewFormalism}). In particular, this 
affects the speed of the propagation of inhomogeneity information 
through coordinate space (as well as some other issues), which 
for now will be calculated with respect to the {\it unperturbed} 
FRW backreaction metric, rather than referencing the perturbed 
metric itself in an explicitly self-consistent manner. 

For an FRW metric with $a(t) = a_{0} (t/t_{0})^{2/3}$, 
the coordinate distance traveled by a null ray in the cosmic 
time span from $t_{1}$ to $t_{2}$ will be $\alpha \equiv 
(c/a_{0}) \int^{t_{2}}_{t_{1}} (t/t_{0})^{-2/3} dt = 
[(3 c / a_{0}) (t_{0})^{2/3} (t_{2}^{1/3} - t_{1}^{1/3})]$. 
Defining $a_{0} \equiv 3 c t_{0} = 2 c / H_{0}$, 
and with $t_{2} \equiv t$, $t_{1} \equiv t_{\mathrm{ret}}$, 
we thus have: 
$\alpha = [(t/t_{0})^{1/3} - (t_{\mathrm{ret}}/t_{0})^{1/3}]$. 
We then turn this into a prescription for computing 
$t_{\mathrm{ret}}$ as a function of $t$ (relative to the 
present time, $t_{0}$) and $\alpha$, as follows:
\begin{equation}
t_{\mathrm{ret}} (t, \alpha) = 
t_{0} [(t/t_{0})^{1/3} - \alpha]^{3} ~ .
\label{EqntRet}
\end{equation}
Similarly, we can determine $\alpha _{\mathrm{max}}$, 
given some initial time $t_\mathrm{Init}$ at which 
structure formation can be reasonably said to have 
started (i.e., 
$\Psi(t \le t_\mathrm{Init}) \equiv 0$):
\begin{equation}
\alpha _{\mathrm{max}} (t, t_\mathrm{Init})
= [(t/t_{0})^{1/3} - (t_\mathrm{Init}/t_{0})^{1/3}] ~ .
\label{EqnalphaMax}
\end{equation}

Now, how the metric at $P$ is affected by a spherical 
shell of material at coordinate radius $\alpha$ depends 
upon the state of clumping there at the appropriate 
retarded time: $\Psi [t_{\mathrm{ret}} (t, \alpha)]$. 
The total effect is then computed by integrating all 
shells from $\alpha = 0$ out to 
$\alpha = \alpha _{\mathrm{max}} (t, t_\mathrm{Init})$; 
but in order to compute the metric perturbation from 
each shell quantitatively, it is first necessary to 
relate this clumping function to an actual physical 
density of material. 

As discussed above, for now consider the $\Psi (t)$ 
function as representing the dimensionless ratio of matter 
which can appropriately be defined as `clumped' at a given 
time, expressed as a fraction of the total physical 
density. Assuming a flat MD cosmology as the initially 
unperturbed state, the total physical density at all times 
will merely be an evolved version of the unperturbed FRW 
critical closure density from early (pre-perturbation) times. 

Recalling Equation~\ref{EqnAngAvgBHMatDomWeakFlat}, 
we have the perturbation term 
$[ R_{\mathrm{Sch}}(t) / r^{\prime} ] = 
\{ (2 G M / c^{2} ) / [r^{\prime} ~ a_{\mathrm{MD}}(t)] \}$, 
with $a_{\mathrm{MD}}(t) = a_{0} (t / t_{0})^{2/3} 
\equiv c [18 t^{2} / H_{0}]^{1/3}$. 
The value of $M$ to use here is given by the 
clumped matter density at coordinate distance $\alpha$ 
times the infinitesimal volume element of the shell. 
The clumped matter density at time $t$, as implied above 
(and still as considered with respect to the unperturbed 
background metric), will equal 
$[ \Psi (t) \rho _{\mathrm{crit}}(t) ]$; and the volume 
element in the integrand representing the effects of that 
shell will be $4 \pi R_{\mathrm{phys}}^{2} d R_{\mathrm{phys}} 
= 4 \pi [a_{\mathrm{MD}}(t) ~ \alpha]^{2} 
[a_{\mathrm{MD}}(t) ~ d \alpha]$. 

Collecting these terms (and letting 
$t^{\prime} \equiv t_{\mathrm{ret}} (t, \alpha)$), 
the integrand will thus be equal to: 
\begin{mathletters}
\begin{eqnarray}
[ R_{\mathrm{Sch}}(t) / r^{\prime} ]
_{r^{\prime} = \alpha \rightarrow (\alpha + d \alpha)} & = & 
\{ (2 G / c^{2} ) ~ d M ~ / 
~ [a_{\mathrm{MD}}(t) ~ \alpha] \} 
\\ 
& = & \{ (2 G / c^{2} ) ~ 
[a_{\mathrm{MD}}(t) ~ \alpha]^{-1} 
~ [ \Psi (t^{\prime}) ~ 
\rho _{\mathrm{crit}}(t) ] 
~ [4 \pi R_{\mathrm{phys}}^{2} d R_{\mathrm{phys}} ] \} 
\\ 
& = & \{ (8 \pi G / c^{2} ) 
~ \Psi (t^{\prime}) 
~ [a_{\mathrm{MD}}(t) ~ \alpha]^{-1} 
~ [\rho _{\mathrm{crit}}(t) ~ 
[a_{\mathrm{MD}}(t)]^{3}] 
~ {[ \alpha^{2} d \alpha} ]\} 
\\ 
& = & \{ (8 \pi G / c^{2} ) 
~ \Psi (t^{\prime}) 
~ a_{\mathrm{MD}}(t)^{-1} 
~ [\rho _{\mathrm{crit}}(t_{0}) ~ a_{0}^{3}] 
~ {[ \alpha d \alpha} ]\} 
\\ 
& = & \{ (8 \pi G / c^{2} ) 
~ \Psi (t^{\prime}) 
~ [(t_{0} / t)^{2/3} ~ (3 c t_{0})^{-1}]
~ \{ [3 H_{0}^{2} / (8 \pi G)]  
~ (3 c t_{0})^{3} \} 
~ {[ \alpha d \alpha} ]\} 
\\ 
& = & \{12 ~ \Psi (t^{\prime}) 
~ [(t_{0} / t)^{2/3}]
~ {[ \alpha d \alpha} ]\} ~ , 
\end{eqnarray}
\label{EqnIintegrandPrelim}
\end{mathletters}
where for simplification we have used 
$H_{0} = (2/3) t_{0}^{-1}$ and the fact that 
$[\rho(t) a(t)^{3}]$ is constant, both true 
for a matter-dominated universe. Note also that 
only $\Psi$ is evaluated at the retarded time, 
$t_{\mathrm{ret}} (t, \alpha)$. The 
{\it strength} of the metric perturbation (at time $t$ 
for point $P$) from a point-like Newtonian perturbation 
embedded in the expansion actually depends upon its 
instantaneous physical distance from $P$ at $t$, 
as is obvious from 
$\Phi = \{ - G M / [a_{\mathrm{MD}}(t) ~ r] \} 
\equiv [- G M / R_{\mathrm{phys}}(t)]$ 
in Equation~\ref{NewtPertSingleClump}. The only 
``relativistic" piece of propagating information 
which is causally delayed is the state of clumping, 
$\Psi [t_{\mathrm{ret}} (t, \alpha)]$, that has 
just then arrived from coordinate distance $\alpha$ 
to observer $P$ at $(r,t)$.

The total integrated metric perturbation function due to 
clumping, $I(t)$, as experienced by a null ray passing 
through point $P$ at $(r,t)$, is thus calculated via the 
causal updating integral: 
\begin{equation}
I(t) = 
\int^{\alpha _{\mathrm{max}} (t, t_\mathrm{Init})}_{0}
\{12 ~ 
\Psi [t_{\mathrm{ret}} (t, \alpha)] 
~ [(t_{0} / t)^{2/3}] \} ~ 
\alpha ~ d \alpha ~ ,
\label{EqnItotIntegration}
\end{equation}
with $I(t)$ implicitly being a function of 
$t_\mathrm{Init}$ (with $I(t) \equiv 0$ for 
$t \leq t_\mathrm{Init}$), as well as of $t_{0}$.

Inserting this result back into the formalism of 
Equation~\ref{EqnAngAvgBHMatDomWeakFlat}, 
the final clumping-perturbed metric that we will use for 
all of our subsequent cosmological calculations becomes: 
\begin{equation}
ds^{2} = 
- c^{2} [ 1 - I(t) ] ~ dt^{2} 
~ + ~ \{ [a_{\mathrm{MD}}(t)]^{2} ~ 
[ 1 + (1/3) I(t) ] \} 
~ \vert d \vec{r} \vert ^{2} ~ . 
\label{EqnFinalBHpertMetric}
\end{equation}

Given this metric, we can define relationships between all 
of the unperturbed FRW (``bare'') cosmological parameters, 
and their corresponding observable (``dressed'') parameters. 
We must first consider cosmological redshifts, where we have: 
\begin{equation}
z^\mathrm{FRW}(t) \equiv 
\frac{a_{\mathrm{MD}}(t_{0})}{a_{\mathrm{MD}}(t)} - 1 
= (t_{0}/t)^{2/3} - 1 ~ ,
\label{EqnDefNzFRW}
\end{equation}
and: 
\begin{equation}
z^\mathrm{Obs}(t) \equiv 
\frac{\sqrt{g_{r r}(t_{0})}}{\sqrt{g_{r r}(t)}} - 1 
= [ \sqrt{\frac{1 + (1/3) I(t_{0})}{1 + (1/3) I(t)}} 
~ (t_{0}/t)^{2/3}] - 1 ~ ,
\label{EqnDefNzObs}
\end{equation}
where quantities without superscripts like 
$t_{0} \equiv t^\mathrm{FRW}_{0}$, etc., will generically refer 
to unperturbed FRW parameters; and where values of observational 
quantities in our backreaction-perturbed formalism will be 
expressly indicated (e.g., $t^\mathrm{Obs}_{0}$). 

Now, in order to calculate observed luminosity distances, 
we must compute the coordinate distance $r$ of a supernova 
going off at coordinate time $t$, for which a light ray 
would be arriving here (at $r = 0$) precisely at $t_{0}$. 
For a null ray and pure inward radial motion we have:
\begin{mathletters}
\begin{eqnarray}
r^\mathrm{FRW}_{\mathrm{SN}}(t) & \equiv & 
\vert r^\mathrm{FRW}(t_{0}) - r^\mathrm{FRW}(t) \vert 
= \int^{t_{0}}_{t}
\{ \sqrt{\frac{g_{t t}(t^{\prime})}{g_{r r}(t^{\prime})}} \} 
~ d t^{\prime} 
\\
& = & 
\int^{t_{0}}_{t}
\{ \frac{c}{a_{\mathrm{MD}}(t^{\prime})} ~ 
\sqrt{\frac{1 - I(t^{\prime})}{1 + (1/3) I(t^{\prime})}} \} 
~ d t^{\prime} 
\\
& = & 
\frac{c}{a_{0}} \int^{t_{0}}_{t}
\{ (t_{0}/{t^{\prime}})^{2/3} 
~ \sqrt{\frac{1 - I(t^{\prime})}{1 + 
(1/3) I(t^{\prime})}} \} 
~ d t^{\prime} 
~ .
\end{eqnarray}
\label{EqnRofTIntegration}
\end{mathletters}

This coordinate distance function can then be converted into 
an expression for the observed luminosity distance by adapting 
the expression from the homogeneous FRW case, for which 
$d_{\mathrm{L}} = [a_{0} ~ r_{\mathrm{SN}} ~ (1 + z)]$, 
as follows:
\begin{mathletters}
\begin{eqnarray}
d_{\mathrm{L}, \mathrm{Pert}}(t) & = & [ a_{0} 
\sqrt{1 + (1/3) I(t_{0})} ] ~ 
r^\mathrm{FRW}_{\mathrm{SN}}(t) ~ [1 + z^\mathrm{Obs}(t)] 
\\
& = & 
\frac{1 + (I_{0}/3)}{\sqrt{1 + [I(t) / 3]}}
~ 
\frac{c ~ t_{0}^{4/3}}{t^{2/3}} 
~ 
\int^{t_{0}}_{t}
\{ (t^{\prime})^{-2/3} 
~ \sqrt{\frac{1 - I(t^{\prime})}{1 + 
[I(t^{\prime}) / 3]}} \} 
~ d t^{\prime} 
\\
& = & 
\frac{1 + (I_{0}/3)}{\sqrt{1 + [I(t_{r}) / 3]}}
~ 
\frac{c ~ t_{0}}{t_{r}^{2/3}} 
~ 
\int^{1}_{t_{r}}
\{ (t^{\prime}_{r})^{-2/3} 
~ \sqrt{\frac{1 - I(t^{\prime}_{r})}{1 + 
[I(t^{\prime}_{r}) / 3]}} \} 
~ d t^{\prime}_{r} 
~ , 
\end{eqnarray}
\label{EqnDlumDefn}
\end{mathletters}
where $I_{0} \equiv I(t_{0})$, and $t_{r}$, 
$t^{\prime}_{r}$ are dimensionless time ratios 
(e.g., $t_{r} \equiv t / t_{0}$), with no change 
to the essential form of $I(t)$ (i.e., 
$I(t) = I(t_{r} \cdot t_{0}) \Rightarrow I(t_{r})$). 

All cosmologically-relevant curves, fits, and parameters can now be 
calculated from the metric (Equation~\ref{EqnFinalBHpertMetric}), and from 
this luminosity distance function (Equation~\ref{EqnDlumDefn}) and its 
derivatives, as investigated in BBI. Important modeled quantities include: 
the residual distance modulus function (with respect to an empty coasting 
universe), $\Delta (m - M)_{\mathrm{Pert}} (z^\mathrm{Obs})$, 
and the quality and probability of its fit, $\chi ^{2} _{\mathrm{Fit}}$ 
and $P_{\mathrm{Fit}}$, to the Type Ia supernova (SNIa) data; 
the observed (versus unperturbed) Hubble Constant, $H^\mathrm{Obs}_{0}$ 
(versus $H^\mathrm{FRW}_{0}$); the physically-measurable age of the 
universe, $t^\mathrm{Obs}_{0}$; the ``true'' value of the total cosmic 
matter density, $\Omega^\mathrm{FRW}_\mathrm{M}$, which determines the 
primordial spatial curvature (i.e., $\Omega^\mathrm{FRW}_\mathrm{M} = 1$ 
for flatness in the pre-perturbed epoch), which our model must relate 
to some measured value of the density, $\Omega^\mathrm{Obs}_\mathrm{M}$, 
for normalization (here and in BBI we use 
$\Omega^\mathrm{Obs}_\mathrm{M} = 0.27$); the observable values 
(defined for $z \rightarrow 0$) of the deceleration parameter 
$q_{0}^{\mathrm{Obs}}$, the effective (total) cosmic equation of state 
$w_{0}^{\mathrm{Obs}}$, and the jerk parameter $j_{0}^{\mathrm{Obs}}$; 
and finally, for comparison with complementary data sets from much 
earlier cosmic epochs, we compute the acoustic scale of the Cosmic 
Microwave Background (CMB) acoustic peaks, $l^\mathrm{Obs}_{\mathrm{A}}$. 

As a technical note, we recall that all evolving quantities in our 
numerical simulation program are calculated as discrete arrays in $t$ 
(and in $z^\mathrm{Obs}(t)$), with a tested pixelization that is fine 
enough for great accuracy in all parameters. Given that the discrete 
version of $d_{\mathrm{L}, \mathrm{Pert}}(z^\mathrm{Obs})$ must be 
differentiated (and evaluated specifically for $z \rightarrow 0$) to 
obtain cosmological parameters, we do so by using the definition of 
the derivative for each pixel, as follows:  
\begin{equation}
[d^{N \prime}_{\mathrm{L}, \mathrm{Pert}}]_{ \{ i \} } 
= \frac{d}{d ~ z^\mathrm{Obs}} 
[d^{(N-1) \prime}_{\mathrm{L}, \mathrm{Pert}}]_{ \{ i \} } 
\equiv \frac{d^{(N-1) \prime}_{\mathrm{L}, 
\mathrm{Pert}}(t_{ \{ i+1 \} }) 
- d^{(N-1) \prime}_{\mathrm{L}, 
\mathrm{Pert}}(t_{ \{ i \} })}
{z^\mathrm{Obs}(t_{ \{ i+1 \} }) 
- z^\mathrm{Obs}(t_{ \{ i \} })}
 ~ , 
\label{EqnPixelDiffDef}
\end{equation}
and then obtain the $z \rightarrow 0$ limit from the last, 
latest-in-time pixel: 
\begin{equation}
[d^{N \prime}_{\mathrm{L}, \mathrm{Pert}}]_{
(z \rightarrow 0, ~ t \rightarrow t_{0})} 
\equiv
[d^{N \prime}_{\mathrm{L}, \mathrm{Pert}}]_{ \{ N_\mathrm{pix} \} } 
 ~ .
\label{EqnFirstEval}
\end{equation}
The cosmological results which we obtain from this procedure appear 
to be robust, with only minor difficulties, as will be discussed 
below; and while the greatest discretization errors occur for 
$j_{0}^{\mathrm{Obs}}$, which requires three differentiations of 
$d_{\mathrm{L}, \mathrm{Pert}}(t)$, virtually all of the results 
which we quote here for $j_{0}^{\mathrm{Obs}}$ should be well 
within $1\%$ in terms of numerical accuracy. 

In order to conduct specific calculations with our formalism, we must 
design a set of physically reasonable clumping evolution functions, 
$\Psi (t)$, to serve as convenient proxies for the combined effects of 
the linear density evolution of early-stage clustering, the nonlinear 
regime and virialization for very dense clumps, and the initial 
development (in many cases triggered by collisions) of entirely new 
clumps with substantial mass. Guided by general cosmological considerations 
and simplicity, in BBI we chose three different classes of time-dependent 
behaviors to examine: $\Psi (t) \propto a(t) \propto t^{2/3}$, 
which is proportional to the evolving contrast of a density variation, 
$\delta \rho / \rho$, in the linear regime \citep{KolbTurner}; 
$\Psi (t) \propto t$, a generally sensible choice depending simply upon 
the amount of time available for clumping; and $\Psi(t) \propto t^{2}$, 
an `accelerating' clumping model which we initially chose as a test case 
to see whether that would possibly help in creating an observed acceleration. 
This last class of models will take on more significance in this paper, 
however, since our results below with recursive nonlinearities will force 
us to regard $\Psi (t)$ as not merely a simple percentage of clumped versus 
unclumped matter, but as a quantity reflecting the details of virialization 
brought to completion (i.e., extremely nonlinear density perturbations) on 
multiple cosmic scales; and appropriately, the density contrast evolution 
for inhomogeneities in the nonlinear regime goes as 
$\delta \rho / \rho \propto a(t)^{n}$ with $n \gtrsim 3$ 
\citep[][p. 322]{KolbTurner} -- that is, $\delta \rho / \rho \propto t^{m}$ 
with $m \equiv (2/3) n \gtrsim 2$ for a matter-dominated universe.

Quantitatively, we defined our three different classes of 
clumping evolution models as follows:
\begin{mathletters}
\begin{eqnarray}
\Psi _{\mathrm{MD}} (t) & \equiv & 
\Psi _{0} ~ 
(\frac{t - t_\mathrm{Init}}{t_{0} - t_\mathrm{Init}})^{2/3}
\\
\Psi _{\mathrm{Lin}} (t) & \equiv & 
\Psi _{0} ~ 
(\frac{t - t_\mathrm{Init}}{t_{0} - t_\mathrm{Init}}) 
\\
\Psi _{\mathrm{Sqr}} (t) & \equiv & 
\Psi _{0} ~ 
(\frac{t - t_\mathrm{Init}}{t_{0} - t_\mathrm{Init}})^{2}
~ . 
\end{eqnarray}
\label{EqnClumpModels}
\end{mathletters}
Here $t_\mathrm{Init}$ (or equivalently, 
$z_\mathrm{Init} = [(t_{0}/t_\mathrm{Init})^{2/3} - 1]$, 
as per Equation~\ref{EqnDefNzFRW}) represents the 
effective beginning of clumping, such that 
$\Psi(t \le t_\mathrm{Init}) \equiv 0$ for all models; and 
$\Psi _{0} \equiv \Psi(t_{0})$ represents the current state of 
clumping today. Each of these models has two physically meaningful 
parameters to vary (besides the universally-optimizable parameter 
for all models, $H^\mathrm{Obs}_{0}$): $\Psi _{0}$, for which 
we selected values by estimating the fractions of Dark and 
baryonic matter that may likely be clumped by $t_{0}$; and 
$z_\mathrm{Init}$, for which we selected values by roughly 
linking the beginning of clumping with the process of 
cosmic reionization. 

In BBI, we evaluated $60$ different models, using 
$\Psi _{0} = (0.78,0.85,0.92,0.96,1.0)$ and 
$z_\mathrm{Init} = (5,10,15,25)$ for each of 
$\Psi _{\mathrm{Lin}} (t)$, $\Psi _{\mathrm{MD}} (t)$, and 
$\Psi _{\mathrm{Sqr}} (t)$. We found that these models produced 
curves which behaved very much like $\Lambda$CDM, with the 
$\Psi _{\mathrm{Sqr}}$ runs looking like flat $\Lambda$CDM with 
$\Omega_{\Lambda} \sim 0.3-0.4$, the $\Psi _{\mathrm{Lin}}$ runs 
looking like $\Omega_{\Lambda} \sim 0.5-0.8$, and the 
$\Psi _{\mathrm{MD}}$ runs looking like $\Omega_{\Lambda} \sim 0.65-0.97$. 

As was expected, due to the increase in the number of perturbing clumps 
with distance (as $r^{2}$), the total perturbative effects were typically 
dominated by the largest distances (and thus the earliest look-back times) 
out to which one could still see significant inhomogeneities; hence 
clumping evolution functions with a more rapid onset of clustering at 
early times would produce a more powerful causal backreaction effect. 
Thus the $\Psi _{\mathrm{MD}}$ models were by far the strongest, while 
the $\Psi _{\mathrm{Sqr}}$ models were all too weak to reproduce the 
observed acceleration, with the $\Psi _{\mathrm{Lin}} (t)$ models 
falling intermediate between the two; and in all cases, the choice of 
an earlier $z_\mathrm{Init}$ led to a stronger overall backreaction. 

All told, roughly a dozen of these models resulted in good fits to the 
standard candle SNIa data, reproducing the observed cosmic acceleration 
essentially as well as the best-fit $\Lambda$CDM Dark Energy model; and 
about half of those dozen models produced especially good cosmological 
parameters, as well. The overall picture was that of a successful attempt 
at achieving an alternative cosmic concordance without Dark Energy. 
Additionally, there emerged the bonus of a testable prediction which could 
distinguish causal backreaction from a Cosmological Constant: a result of 
$j_{0}^{\mathrm{Obs}} \gg 1$ for all of our models which produced good fits, 
in contrast to the mandatory value of $j_{0}^{\mathrm{Obs}} = 1$ for all 
flat $\Lambda$CDM models, regardless of the value of $\Omega_{\Lambda}$. 

With those results in hand, the next concern was to determine how the effects 
of recursive nonlinearities might modify these outcomes, once incorporated into 
our numerical models. Our initial expectation was that the overall Hubble curves 
of each modeled cosmology would likely not change significantly, except perhaps 
at the end, as $z \rightarrow 0$; a result that would not alter the SNIa fitting 
results much, but which could have a measurable effect upon the cosmological 
parameters evaluated at $t_{0}$ -- particularly upon our 
`paradigm falsifiability' parameter, $j_{0}^{\mathrm{Obs}}$. The actual outcome, 
however, is that the proper treatment of recursive nonlinearities ends up 
altering the situation to a great degree, requiring us to re-evaluate (see 
Section~\ref{SecRecNonlinResults}) the ultimate implications of causal 
backreaction; although it does still remain true that an alternative concordance 
can be successfully achieved with it. First, however, in the following 
subsection we give the precise details explaining how recursive nonlinearities 
for causal backreaction are technically implemented in our numerical simulations.

\subsection{\label{SubSecNewFormalism}The New Formalism: Incorporating 
                      Recursive Nonlinearities into Causal Backreaction} 

Though in practice necessitating a complete re-write of our numerical 
simulation program, the essential features of recursive nonlinearities 
require just two fundamental changes. 

The first change involves the rate at which clustering information 
can propagate through the inhomogeneity-perturbed universe, as depicted 
earlier in Figure~\ref{FigSNRayTraceInts}. The result is that 
$t_{\mathrm{ret}} (t, \alpha)$ and 
$\alpha _{\mathrm{max}} (t, t_\mathrm{Init})$ are altered from their 
simple functional forms (as per Equations~\ref{EqntRet},\ref{EqnalphaMax}) 
for a matter-dominated FRW universe, and now depend 
upon the metric perturbation potential function, $I(t)$. But as we see 
from Equation~\ref{EqnItotIntegration}, $I(t)$ is itself calculated 
via an integral depending upon $t_{\mathrm{ret}} (t, \alpha)$ and 
$\alpha _{\mathrm{max}} (t, t_\mathrm{Init})$. Hence $I(t)$ is now 
defined in terms of $I(t)$, in a fundamentally recursive fashion. 

Second, the extra spatial volume created by nonzero $I(t)$ will 
have a dilution effect upon the strength of the perturbation induced 
by any (now farther-away) clumped mass upon the metric at our 
observation point. Examining Equation~\ref{EqnIintegrandPrelim} 
(specifically \ref{EqnIintegrandPrelim}a), we recall that the 
perturbation term for each spherical shell of clumped matter 
is given by $[ R_{\mathrm{Sch}}(t) / r^{\prime} ]
_{r^{\prime} = \alpha \rightarrow (\alpha + d \alpha)} = 
\{ (2 G / c^{2} ) ~ d M ~ / 
~ [a_{\mathrm{MD}}(t) ~ \alpha] \}$. Now, the differential mass 
element in the shell, $d M$, does not actually change (and therefore 
requires no correction factor), since the dilution of its mass 
density is precisely offset by its expanded volume. But what 
{\it does} change is the effective distance of those perturbing clumps 
from the observation point. Recalling that the strength of a 
Newtonian perturbation embedded in an expanding universe depends simply 
upon its instantaneous physical distance, we see that the denominator 
$[a_{\mathrm{MD}}(t) ~ \alpha]$ must now be replaced by the term: 
$[\sqrt{g_{r r}} ~ \alpha] 
= [a_{\mathrm{MD}}(t) ~ \sqrt{1 + (1/3) I(t)} ~ \alpha]$, which 
in the end just puts a factor of $\sqrt{1 + (1/3) I(t)}$ into 
the denominator of the integral for $I(t)$. 

Therefore, the modified formula (replacing 
Equation~\ref{EqnItotIntegration}) for calculating the 
metric perturbation function $I(t)$ with the incorporation 
of recursive nonlinearities (``RNL''), is given as: 
\begin{equation}
I^{\mathrm{RNL}}(t) = 
\int^{\alpha _{\mathrm{max}} (t, t_\mathrm{Init}, I)}_{0}
\frac{\{12 ~ 
\Psi [t_{\mathrm{ret}} (t, \alpha, I)] 
~ [(t_{0} / t)^{2/3}] \} ~ 
\alpha}{\sqrt{1 + (I/3)}} ~ d \alpha ~ , 
\label{EqnItotRNLintegration1}
\end{equation}
where the term ``$I$'' inside the integral on the right-hand side 
represents the actual function, $I^{\mathrm{RNL}}(t)$, itself. But 
since the denominator term is in fact independent of the integration 
variable $\alpha$, we can remove it from the integral and bring it to 
the left-hand side of the equation. The metric perturbation function can 
therefore be (numerically) solved for any given $t$ as the solution of: 
\begin{equation}
I^{\mathrm{RNL}}(t) ~ \sqrt{1 + [ I^{\mathrm{RNL}}(t) /3]} = 
\int^{\alpha _{\mathrm{max}} (t, t_\mathrm{Init}, I)}_{0}
\{12 ~ 
\Psi [t_{\mathrm{ret}} (t, \alpha, I)] 
~ [(t_{0} / t)^{2/3}] \} ~ 
\alpha ~ d \alpha ~ . 
\label{EqnItotRNLintegration2}
\end{equation}

To evaluate the expressions $t_{\mathrm{ret}} (t, \alpha, I)$ and 
$\alpha _{\mathrm{max}} (t, t_\mathrm{Init}, I)$ in this above integral, 
we utilize (analogously with the discussion preceding 
Equations~\ref{EqntRet},\ref{EqnalphaMax}): 
\begin{equation}
\alpha (T, T_{\mathrm{ret}}, I) = 
\frac{1}{3} \int^{T}_{T_{\mathrm{ret}}} 
\frac{\sqrt{1 - I^{\mathrm{RNL}}(T^{\prime})}}{\sqrt{1 
+ [I^{\mathrm{RNL}}(T^{\prime}) / 3]}} 
\frac{d T^{\prime}}{(T^{\prime})^{2/3}} ~ , 
\label{EqnAlphaRNLintegration2}
\end{equation}
with $T \equiv (t/t_{0})$ and 
$T_{\mathrm{ret}} \equiv (t_{\mathrm{ret}}/t_{0})$. This expression 
for $\alpha (T, T_{\mathrm{ret}}, I)$ could in theory be inverted to 
produce $T_{\mathrm{ret}} (T, \alpha, I)$; and in addition, we have 
$\alpha _{\mathrm{max}} = \alpha (T, T_{\mathrm{Init}}, I)$. 

In practice, we perform these recursively-defined `integrals' by 
utilizing discrete arrays in cosmic coordinate time with some large number 
of pixels covering the range from $T_{\mathrm{Init}}$ to $T_{0} \equiv 1$, 
for which the later pixels are calculated in terms of the earlier pixels. 
Beginning with the first pixel at $T_{\mathrm{Init}}$, we thus have the 
recipe (with $N_\mathrm{pix}$ pixels, and 
$\Delta T = [(T_{0} - T_{\mathrm{Init}})/( N_\mathrm{pix} - 1)]$): 
\begin{mathletters}
\begin{eqnarray}
\alpha _{\mathrm{max},1} & = & I^{\mathrm{RNL}}_{1} \equiv 0 ~, 
\\
\alpha _{\mathrm{max},2} & = & 
\frac{1}{3} \frac{\Delta T}{T_{\mathrm{Init}}^{2/3}} ~, 
~ I^{\mathrm{RNL}}_{2} = 0 ~, 
\end{eqnarray}
\label{EqnRNLdiscreteGridLoopA}
\end{mathletters}
and then, for $i = \{ 3, N_\mathrm{pix} \}$: 
\begin{mathletters}
\begin{eqnarray}
& \alpha _{\mathrm{max},i} & = \alpha _{\mathrm{max},(i-1)} 
+ \{ \frac{1}{3} 
\frac{\sqrt{1 - I^{\mathrm{RNL}}_{(i-1)}}}{\sqrt{1+[I^{\mathrm{RNL}}_{(i-1)}/3]}} 
\frac{\Delta T}{[T_{(i-1)}]^{2/3}} \} ~,  
\\
& X^{\mathrm{RNL}}_{i} & = \frac{12}{T_{i}^{2/3}} 
~ \sum_{k = \{1,(i-2)\}} \{ \Psi[T_{(i-k)}] 
[\alpha _{\mathrm{max},i} - \alpha _{\mathrm{max},(i-k)}] 
[\alpha _{\mathrm{max},(i+1-k)} - \alpha _{\mathrm{max},(i-k)}] \} ~,  
\\
& I^{\mathrm{RNL}}_{i} & \sqrt{1 + [ I^{\mathrm{RNL}}_{i} /3]} 
= X^{\mathrm{RNL}}_{i} 
~. 
\end{eqnarray}
\label{EqnRNLdiscreteGridLoopB}
\end{mathletters}

The result of this iterative loop is the discrete array 
$\{ I^{\mathrm{RNL}}_{i} \}$, which serves as $I^{\mathrm{RNL}}(t)$ 
for all cosmological calculations, as described above. As a test, we 
have verified that removing the effects of recursive nonlinearities 
(essentially adjusting to unity all terms inside the radical signs in 
Equations~\ref{EqnRNLdiscreteGridLoopB}a,c) succeeds in reproducing the 
results of our old formalism to a great degree of accuracy; and for our 
new model simulations with the recursive nonlinearities included, we 
have checked to make sure that the full suite of results with this 
new formalism appears quite sensible and consistent in all cases. 

We have also tested the pixelization for precision of the output results, 
finding that runs with $\sim$$1000$ pixels are sufficient for quick model 
parameter optimization searches, producing all cosmological results to 
within a couple percent of their `true' values (i.e., the values to which 
the parameters asymptote for much larger pixelizations); and that going 
to $\sim$$5000-10000$ pixels yields cosmological output results that are 
stable to within a small fraction of a percent. (Though going all the way 
to $\sim$$25000$ pixels actually leads to discretization problems due 
to round-off error, which causes problems for higher derivatives of the 
luminosity distance function, leading to instability in the value of 
$j_{0}^{\mathrm{Obs}}$; and so we avoid pixelizations this large.) Our 
high-precision output cosmological parameters -- i.e., all numerical 
results other than those from optimization searches -- which we will 
quote in Section~\ref{SecRecNonlinResults}, below, have all been 
obtained from runs with $\sim$$5000-10000$ pixels (not counting the 
further addition of a smaller number ($\sim$$1000$) of extra pixels 
used for integrating the Hubble curves out to the computationally 
simpler region past $z_\mathrm{Init}$, before the simulated onset 
of clumping).

\section{\label{SecRecNonlinResults}RESULTS WITH RECURSIVE NONLINEARITIES} 

\subsection{\label{SubUnityClumpWeak}Weakened Backreaction Effects: 
Difficulties with ``Clumping-Saturated'' Models} 

\begin{figure}
\begin{center}
\includegraphics[scale=1.0]{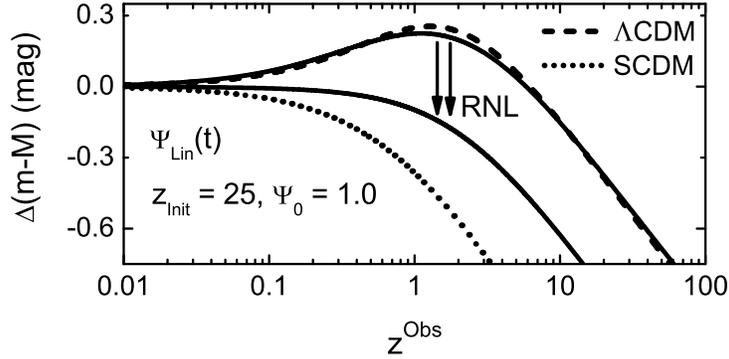}
\end{center}
\caption{Residual Hubble diagrams for the causal 
backreaction model $\Psi _{\mathrm{Lin}} (t)$ with 
$(z_\mathrm{Init},\Psi _{0}) = (25,1.0)$, where the 
upper solid line represents the old version of the 
simulated cosmology without recursive nonlinearities 
(RNL), and the lower solid line represents the 
new version with RNL. Shown along with them for 
comparison (broken lines) are the flat SCDM and 
(Union1-best-fit, $\Omega _{\Lambda} = 0.713$) 
Concordance $\Lambda{\mathrm{CDM}}$ cosmologies.}
\label{FigStrCL1WeakPlot}
\end{figure}

As mentioned previously, the incorporation of recursive nonlinearities 
(henceforth RNL) does not merely affect the late-time behavior as 
$z \rightarrow 0$, but in fact exerts a profound damping influence 
upon the entire process of causal backreaction, weakening the overall 
effect for a given set of model input parameters. As an example, we 
consider the impact of RNL upon one of the best-fitting models from 
BBI, $\Psi _{\mathrm{Lin}} (t)$ with $z_\mathrm{Init} = 25$ and 
$\Psi _{0} = 1.0$. As depicted in Figure~\ref{FigStrCL1WeakPlot}, we 
see that adjusting the simulations to include RNL greatly reduces the 
apparent acceleration effect -- no longer quite even achieving an 
`acceleration', in fact, since $q_{0}^{\mathrm{Obs}} = 0.058 > 0$. 

For ease of comparison to the results from BBI, in this paper we 
will still primarily use the SNIa data from the SCP Union1 supernova 
compilation \citep{KowalRubinSCPunion} for conducting fits of our new 
models with RNL. As would be expected from Figure~\ref{FigStrCL1WeakPlot}, 
the new fit to these SNIa data (not shown here), even after 
re-optimization with respect to $H^\mathrm{Obs}_{0}$, is much worse 
in this modified model, with its `chi-squared' value -- for $307$ 
SNIa minus $3$ model parameters yielding $304$ degrees of freedom 
-- increasing from $\chi^{2}_{\mathrm{Fit}} = 312.1$ without RNL, 
to $\chi^{2}_{\mathrm{Fit}} = 410.1$ with it (where for comparison, 
$\chi^{2}_{\mathrm{Fit}} = 311.9$ for best-fit flat $\Lambda$CDM). 
Visually speaking, though this causal backreaction model with RNL 
does at least manage to clearly separate itself from the strongly 
decelerating behavior of matter-only flat SCDM, it no longer 
matches best-fit $\Lambda$CDM but now falls significantly short of 
it, lying somewhere in the middle between SCDM and $\Lambda$CDM; 
and correspondingly, the output cosmological parameters -- though 
still being substantially modified from those for the unperturbed 
case without causal backreaction -- are now far less reflective of 
those required for a successful concordance, as would indeed be 
expected from a cosmological model producing too weak of 
an `acceleration' effect.

\placefigure{FigStrCL1WeakPlot}

\begin{figure}
\begin{center}
\includegraphics[scale=1.0]{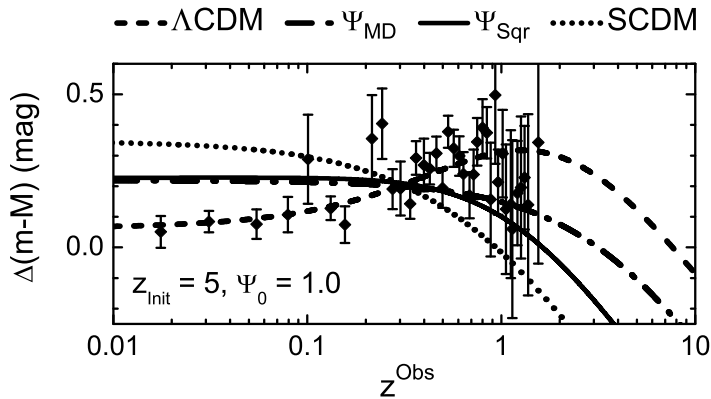}
\end{center}
\caption{Residual Hubble diagrams for the two `best' 
$\Psi _{0} \le 1$ runs (as described in the text), 
selected from the new simulations with RNL but with the 
choice of model input parameters restricted to those 
used in BBI; specifically, these $\Psi _{\mathrm{MD}}$ 
and $\Psi _{\mathrm{Sqr}}$ curves both have 
$(z_\mathrm{Init},\Psi _{0}) = (5,1.0)$. Also 
plotted here are the Union1-best-fit flat 
SCDM and Concordance $\Lambda{\mathrm{CDM}}$ 
($\Omega _{\Lambda} = 0.713$) cosmologies. Shown along 
with these curves are the SCP Union1 SNIa data, here 
binned and averaged for visual clarity 
(bin size $\Delta \mathrm{Log}_{10} [ 1 + z ] = 0.01$). 
Each theoretical model is individually optimized 
in $H^\mathrm{Obs}_{0}$ to minimize its 
$\chi^{2}_{\mathrm{Fit}}$ with respect to the Union1 SNIa data 
set; and for simplicity, instead of moving the SNIa data up or 
down for each different optimized $H^\mathrm{Obs}_{0}$ value, 
the optimization is depicted here by plotting the residual 
Hubble diagram of the SNIa data versus a coasting universe 
of a single, fixed Hubble constant 
($H^\mathrm{Obs}_{0} = 72 ~ \mathrm{km} ~ \mathrm{s}^{-1} 
\mathrm{Mpc}^{-1}$), 
and then displacing each theoretical curve vertically, 
relative to the SNIa data, as appropriate for each fit.}
\label{FigSqrMDCL1optH0WeakPlot}
\end{figure}

Within the set of $60$ model input parameter choices from BBI (all 
with $\Psi _{0} \le 1$), the `best' models now, with RNL included, are 
$\Psi _{\mathrm{MD}}$ with $(z_\mathrm{Init},\Psi _{0}) = (5,1.0)$, 
which fits the SNIa most successfully with 
$\chi^{2}_{\mathrm{Fit}} = 397.4$; and $\Psi _{\mathrm{Sqr}}$, also 
with $(z_\mathrm{Init},\Psi _{0}) = (5,1.0)$, which despite yielding 
a worse overall fit ($\chi^{2}_{\mathrm{Fit}} = 421.3$) because of 
too little clumping at early times, manages to produce the largest 
late-time backreaction effect, achieving an actual `acceleration' with 
$q_{0}^{\mathrm{Obs}} = -0.026 < 0$. These two models are plotted 
in Figure~\ref{FigSqrMDCL1optH0WeakPlot}, shown along with matter-only 
flat SCDM and $\Lambda$CDM, where all models are now shown in 
the figure as optimized with respect to $H^\mathrm{Obs}_{0}$ for the 
Union1 SNIa data set.

\placefigure{FigSqrMDCL1optH0WeakPlot}

One of the most important changes in causal backreaction due 
to RNL, is the slowdown that it causes for inhomogeneity information 
propagating in from great cosmological distances. This strongly damps 
the backreaction effects due to perturbations at the outer edges of an 
observer's causal inhomogeneity horizon, which despite being the most 
distant ones are nevertheless very important for backreaction because 
they include the largest spherical shells of inhomogeneous matter, 
containing the greatest incremental mass (per shell) of perturbations 
overall. 

Besides simply weakening the total backreaction effect, RNL acts 
specifically to inhibit the contributions from older perturbations 
generated during relatively early clumping epochs, due to a reduction 
of the coordinate distance (and thus of the volume of clumped matter 
now enclosed by the sphere) out to some given propagative look-back 
time; an inhibition that is particularly strong if the perturbation 
potential $I(t)$ has grown too close to unity very early on. One 
important potential result is that this may significantly lessen the 
ongoing effects of causal backreaction into the far future -- due in 
the pre-RNL formalism to eternally-expanding observational horizons 
-- an issue related to the ultimate possible fates of the universe, 
to be discussed further in Section~\ref{SubNewFuture}, below. But 
our immediate concern here is to note that RNL makes later (i.e., 
more recently generated) clumping far more effective than earlier 
clumping in terms of generating a late-time acceleration. Hence in 
our runs here with RNL, we obtain the strongest apparent acceleration 
effects with the $\Psi _{\mathrm{Sqr}}$ models, and the weakest with 
the $\Psi _{\mathrm{MD}}$ models; also, choosing smaller (i.e., later) 
values of $z_\mathrm{Init}$ always makes $q_{0}^{\mathrm{Obs}}$ more 
`accelerative' (smaller or more negative), as well. This is all very 
different from the results in BBI, where the $\Psi _{\mathrm{Sqr}}$ 
functions were by far the weakest models at producing an apparent 
acceleration; and where the $\Psi _{\mathrm{MD}}$ models were so 
strong that small or mid-range values of $z_\mathrm{Init}$ were 
necessary to obtain good fits, since those models with the beginning 
of clumping occurring significantly {\it earlier} than 
$z_\mathrm{Init} \sim 5$ led to an actual overkill of acceleration. 

This behavior of causal backreaction with RNL suggests that we may 
be able to get a stronger accelerative effect by using later and 
later (i.e., smaller) values of $z_\mathrm{Init}$; and indeed, 
going from $z_\mathrm{Init} = 5$ to $z_\mathrm{Init} = 2$ for the 
$\Psi _{\mathrm{Sqr}}$ model with $\Psi _{0} = 1.0$ does strengthen 
the observed acceleration from $q_{0}^{\mathrm{Obs}} = -0.026$ to 
$q_{0}^{\mathrm{Obs}} = -0.043$. But this type of procedure does not 
save these models as a mechanism for explaining the apparent 
acceleration, due to several obvious problems. First of all, the 
total amount of generated acceleration is still far too small, and 
thus the fits to the SNIa data remain extremely poor (and in fact, 
$\chi^{2}_{\mathrm{Fit}}$ gets {\it worse} for the 
$\Psi _{\mathrm{Sqr}}$ runs with later $z_\mathrm{Init}$). Besides 
such problems with the SNIa fits, the output cosmological parameters 
are also unacceptable for such runs, with quantities like 
$t^\mathrm{Obs}_{0}$ and $\Omega^\mathrm{FRW}_\mathrm{M}$ growing 
far too small with decreasing $z_\mathrm{Init}$, thus spoiling any 
efforts at achieving an alternative cosmic concordance. Lastly, 
such input parameters do not even make good physical sense -- can 
we seriously estimate the effective onset of cosmic clustering as 
beginning at $z_\mathrm{Init} \sim 2$, or even later? Clearly, we 
cannot achieve our goals of reproducing the observed cosmic 
evolution with causal backreaction in the presence of RNL while 
stuck within the confines of models with $\Psi _{0} \le 1.0$. 

These difficulties lead us to the question of what may be wrong with 
our formalism; a question to which there are several possible answers. 

First of all, it is certainly possible that causal backreaction is 
not the mechanism fundamentally responsible for the observed cosmic 
acceleration. Even so, given that these effects are (as we argue) 
real and significant -- the cosmic expansion is clearly different 
than it would be without them, as we have seen above in 
Figures~\ref{FigStrCL1WeakPlot} and \ref{FigSqrMDCL1optH0WeakPlot}, 
witnessing the clear separation between our models and unperturbed 
SCDM -- even a cosmology actually dominated by some form of 
Dark Energy would be strongly modified by causal backreaction, 
leading to misconceptions about the evolution of the universe 
(and about the nature of that Dark Energy, itself) were these effects 
not taken into account. Alternatively, causal backreaction may be 
part of a `combination' solution for explaining the apparent 
acceleration -- in conjunction, perhaps, with observational effects 
(recall Section~\ref{SubSecOldApprox}) such as lensing or other 
inhomogeneity-induced perturbations to the luminosity distance 
function $d_{\mathrm{L}} (z)$  -- such that the {\it total} result 
from all perturbative effects, combined, may be sufficient to remove 
the need for Dark Energy. 

Another possibility is that our fundamental physical approach towards 
causal backreaction is far too oversimplified to properly model the 
cosmic evolution, just as we have argued was true for the standard 
(non-causal) method traditionally used for estimating backreaction. 
There are many possible flaws with our approach -- for example, 
perhaps the linearized McVittie solution (from which we calculated 
our perturbative metric in Section~\ref{SubSecOldFormalism}) is 
inappropriate, given that the total mass of the perturbation in 
that metric is fixed as constant in time \citep{KaloKlebanBHinFRW}, 
in contrast to the tendency of overdensities in the real universe 
to keep increasing in mass due to incoming mass flows. (Though our 
growing $\Psi(t)$ functions are of course intended as an alternative 
way of accounting for this.) On the other hand, it is quite possible 
that the complex processes of cosmic structure formation are too 
dynamically rich to be modeled accurately by a Newtonianly-perturbed 
metric containing just a single perturbation potential function. 
The very natures of our $\Psi(t)$ functions are obviously 
oversimplified -- not accounting at all, for example, for the 
differences between the simple linear evolution of matter 
overdensities, and the far more complex process of nonlinear density 
perturbation evolution -- the latter including subsidiary effects 
such as virialization and stabilization (the detailed time-dependence 
of which may be quite important, though our formalism completely 
neglects it), as well as ``gastrophysics'' feedback (e.g., energy 
injection from star formation and from the supernovae, themselves) 
resulting in baryon shock heating, and so on. Considering such 
complications, it is possible that no simple functional form of 
{\it any} kind for $\Psi(t)$ -- and for that matter, no simple 
numbers for $z_\mathrm{Init}$ or $\Psi _{0}$ -- may suffice for 
a realistic description of the physics. Perhaps nothing would be 
accurate short of a fully-detailed cosmic simulation program, 
calculating a realistically self-consistent representation of 
the intertwined behaviors of cosmic structure formation and 
astrophysical feedback processes as they evolve together 
over time. 

A different, possibly significant flaw may be our complete 
neglect of {\it gravitational} nonlinearities; after all, values 
for the `Newtonian' perturbation potential in the range of 
$I_{0} \sim 0.2 - 0.6$ are certainly not negligible, and may 
imply higher-order gravitational terms from general relativity 
that would contribute significant backreaction effects for any 
reasonable model of the evolution of cosmic structure. Leaving 
out such nonlinear gravitational terms may imply that our 
calculations here are not serving as an `estimate' of the 
overall causal backreaction effect, but rather as a 
{\it lower limit} to it. 

Yet another possibility is that the greatly simplifying assumption 
of a smoothly-inhomogeneous universe is incompatible with a 
proper treatment of recursive nonlinearities; after all, 
inhomogeneities massive on a truly cosmological scale, such as galaxy 
clusters, are not smoothly spread in angle, but are concentrated at 
particular locations on the sky. As such, RNL effects would slow down 
the propagation of inhomogeneity information more strongly along some 
directions than along others; and so perturbative effects from new 
clumps, located in new directions, may be able to `slip in' with 
less delay than would inhomogeneity information coming in from the 
same directions on the sky as pre-existing clusters that have already 
imposed their causal backreaction effects upon us. (A kind of 
{\it cosmic} ``channeling'' effect, so to speak.) But quantitatively 
significant or not, it is likely that such issues cannot be properly 
evaluated without the development of a fully-3D simulation program 
for causal backreaction, as discussed previously in 
Section~\ref{SubSecOldApprox}. 

On the other hand, the main difficulty here may really be much 
simpler: the basic physical assumptions and methods of our formalism 
might actually be capable of modeling causal backreaction effects 
well enough, with the problem simply being inadequate choices for 
our (heuristically-adopted) clumping evolution models and model input 
parameters. Given the above results with RNL, it is clear that simply 
tweaking the detailed time-dependence of the $\Psi (t)$ functions 
would not be enough to obtain good data fits; but despite such 
difficulties, there is actually one very simple thing that we 
{\it can} do, which may extend the applicability of our causal 
backreaction simulation formalism, while also dealing with some 
very real physical effects in an empirical yet useful way. From 
the results in this section, it is clear that we cannot achieve a 
new cosmic concordance with what one might call ``clumping-saturated'' 
models that are ultimately limited in strength to values of $\Psi _{0}$ 
bounded above by unity. But what about models with $\Psi _{0} > 1$? 
Such models would clearly lead to much stronger backreaction effects 
(with no pre-determined upper bound, obviously), and could perhaps 
include models that explain all cosmological observations, 
re-establishing an alternative concordance for causal backreaction 
even in spite of the damping effects of RNL. Given our basic definition 
of $\Psi _{0}$ as ``the fraction of cosmic matter in the clumped, 
self-stabilized state'', one may ask whether it is ever physically 
appropriate to use values of $\Psi _{0}$ in excess of unity. In the 
upcoming section, we will argue that the answer is {\it yes}, that 
it is entirely appropriate -- and in fact, both useful and 
necessary -- for a valid phenomenological description 
of {\it hierarchical clustering} in the universe.

\subsection{\label{SubSuperUnityClumpStrong}Strengthened Backreaction Effects: 
Regaining an Observed Acceleration with Hierarchical, Nonlinear Clustering} 

The convenient but ad-hoc representation of the state of cosmic clumping at 
some given time $t$ with just a single number -- $\Psi (t)$ -- is clearly 
an extreme simplification, as discussed above. Even aside from our neglect 
of the detailed time dependence of virialization processes by having this 
function simply represent the clumped matter fraction over time, 
there is also the fact that large-scale structure is not a featureless blob: 
the `clumped' state is not a simple binary state, `on' or `off'. Most notably, 
clustering in the universe is not structure-free, but instead exhibits a 
roughly discrete form of hierarchical clustering, with identifiable structures 
existing on a few well-defined scales -- for example, galaxies existing as 
virialized groupings of stars, galactic clusters as virialized groupings of 
galaxies, and so on. This recognizable substructure will obviously affect the 
resulting causal backreaction that is induced, and so the question becomes 
how to best (and most simply) estimate the effects of such substructure. 

One conclusion which we can immediately draw is that clustering with multiple 
levels of substructure will exert {\it more} of a total backreaction effect 
than would a simpler form of structure without them. From elementary physical 
considerations, it is clear that the total gravitational potential energy 
contained within a bound collection of individually-condensed clumps will be 
larger in magnitude (i.e., more negative) than that within a bound volume 
of featureless dust. By the Virial Theorem, the presence of a greater amount of 
(negative) potential energy requires the counterbalance of more kinetic energy, 
which implies more backreaction-causing vorticity and velocity dispersion. 
Thus the existence of more (stable) substructure implies that more induced 
causal backreaction was generated during the process of stabilizing it. 

But one must ask, how much more? Each distinct level of structure actually 
involves the {\it same} amount of mass, when considering it over a fixed 
cosmic volume -- whether one considers the individual virializations of the 
tens or hundreds of galaxy clusters in a single supercluster, or of the tens 
or hundreds of {\it thousands} of individual galaxies in that same supercluster, 
or of the trillions of stellar-mass bodies in that supercluster -- and in 
virializing the same amount of mass, it seems reasonable to expect the same 
amount of volume-generating velocity dispersion to occur. Thus we claim that 
an effective phenomenological way of estimating the total backreaction effect 
that has been generated by a stabilized volume of mass with multiple levels 
of internal substructure on different length scales, is by {\it counting the 
effective number of levels of structure} that have attained an essentially 
complete virialization. Astrophysically realistic structure formation 
therefore provides a natural way in which $\Psi (t)$ should be expected to 
not only reach but to eventually exceed unity, and then to even keep growing 
(without any fundamental theoretical limit) as long as there are new levels 
of structure to be brought into a self-stabilized state. 

To avoid giving ourselves an infinite amount of leeway in searching for an 
alternative concordance, however, it is important to place bounds upon the 
`reasonable' values of $\Psi (t _{0}) \equiv \Psi _{0}$ that one should expect 
to find when using causal backreaction to explain the apparent acceleration. 
The essence of the question is simple: ``How many levels?'' In particular, 
we need only consider those typical length scales which became 
virialized during or not too long before the general cosmic time of the 
observed acceleration (i.e., $z \lesssim$ few), since (as will be shown 
in Section~\ref{SubEarlySatClumping}) RNL sharply limits the ongoing 
influence of causal backreaction effects from relatively early, 
`outdated' virializations ({\it especially} when they are strong). 

It seems safe to include at least two scales in our $\Psi _{0}$ estimate: 
that of individual galaxies, and that of galaxy clusters, implying a 
$\Psi _{0}$ of at least $2$. Going below these classes of structures in 
length scale, there are of course a multitude of individual star clusters 
and solar systems (supported by the vorticity in their orbital motions) that 
have not too long ago undergone (or are still undergoing) relaxation; and 
even below that in scale, there are continual generations of newly-forming 
solid objects supported by internal body forces (which going beyond the 
perfect-fluid approximation, also contribute positive terms to the 
Raychaudhuri equation). Below this, of course, there is infinite regress; 
though at some point it becomes absurd to consider ever-smaller structures, 
and significant quantities of new self-stabilization is not happening at 
very small scales in any case. It is difficult to precisely estimate the 
total effective causal backreaction feedback that is induced by structures 
at sub-galaxy scales; but for our purposes, let us consider it a 
`significant fraction' of $1$. 

At much higher scales, there are groupings of galactic clusters, leading to 
superclusters, and the largest known structures in general. Superclusters 
are too large to be gravitationally bound yet (and certainly were not bound 
at high redshifts, back around the onset of the apparent acceleration), but are 
experiencing relaxation processes that are underway during the current epoch. 
Once again lacking a precise estimate, we hypothesize that the contributions 
by stabilized structure formation to backreaction from scales larger than that 
of typical galaxy clusters, in terms of increasing the effective value of 
$\Psi _{0}$, may amount to a small but non-negligible fraction of $1$. 

All told, reasonable values of the present-day clumping evolution parameter 
for causal backreaction, when one considers hierarchical clustering, may be 
within the range $\Psi _{0} \sim 2 - 3$; or really pushing it, perhaps within 
$\Psi _{0} \sim 2 - 4$. Allowing ourselves such an enhanced phase space for 
$\Psi _{0}$, the question then becomes whether or not we can re-establish 
causal backreaction, now with RNL included, as a successful 
(and `concordant') explanation of the observed acceleration. 

For each of the $\Psi _{\mathrm{MD}}$, $\Psi _{\mathrm{Lin}}$, and 
$\Psi _{\mathrm{Sqr}}$ clumping evolution functions, different values 
of $z_\mathrm{Init}$ can be explored, where the highest reasonable value 
(say, $z_\mathrm{Init} \sim 25$) is set to represent a time shortly before 
the onset of cosmic reionization; and where the smallest reasonable value 
(say, $z_\mathrm{Init} \sim 2$) would represent the time leading up to 
serious and intense virialization, signaled by such events as the supernovae 
themselves (noting that the highest-redshift SNIa in compilations like those 
of the SCP Union collaboration tend to be at $z \sim 1.5$ or so). Then for 
each value of $z_\mathrm{Init}$ that we adopt, a search is done over 
$\Psi _{0}$ to find which value achieves a best-fit by minimizing 
$\chi^{2}_{\mathrm{Fit}}$. 

Finding a value of $\Psi _{0}$ which minimizes $\chi^{2}_{\mathrm{Fit}}$ for 
any given $z_\mathrm{Init}$ can always be done -- and often with good results, 
where the $\chi^{2}_{\mathrm{Fit}}$ value is almost as low (and sometimes even 
lower) than that for the best-fit $\Lambda$CDM model. But there is more to the 
true task of `optimization' than this, since besides achieving a good fit to 
the SNIa data, one must also match the observed values of a variety of 
complementary cosmological parameters in order to produce a proper alternative 
concordance. Achieving all of these goals simultaneously is more challenging; 
and so to conveniently measure our ability in this regard, we define what can 
roughly be considered an ``Average Percent Deviation from Concordance'' 
parameter, as follows: 
\begin{eqnarray}
\mathrm{Avg. ~ \% ~ Dev.} & = & \frac{1}{6} \{ 
\frac{\vert \Omega^\mathrm{FRW}_\mathrm{M} - 1 \vert}{1} 
+ \frac{\vert H^\mathrm{Obs}_{0} - 69.96 \vert}{69.96} 
+ \frac{\vert t^\mathrm{Obs}_{0} - 13.64 \vert}{13.64} 
\nonumber 
\\ 
& + & \frac{\vert w^\mathrm{Obs}_{0} - (-0.713) \vert}{\vert -0.713 \vert} 
+ \frac{\vert l^\mathrm{Obs}_{\mathrm{A}} - 285.4 \vert}{285.4} 
+ \frac{(\chi^{2}_{\mathrm{Fit}} - 311.9)}{311.9} 
\} \times 100 \% ~ , 
\label{EqnAvgPctConcordDev}
\end{eqnarray}
where all of the numerical values used above to represent the 
`Concordance' values are reference numbers obtained from the 
Union1-best-fit flat $\Lambda$CDM model (i.e., 
$\Omega _{\Lambda} = 0.713$, $H_{0} = 69.96 ~ \mathrm{km} ~ 
\mathrm{s}^{-1} \mathrm{Mpc}^{-1}$, with the approximation 
of zero cosmological radiation density), and all quantities are 
expressed here in typical units (e.g., GYr for $t^\mathrm{Obs}_{0}$). 

An optimization search over the range of reasonable input parameters 
for our three cluster evolution functions, using quick simulation runs 
(i.e., 1000 pixels for $t \ge t_\mathrm{Init}$), produces a number of 
models that provide both good SNIa fits and good output cosmological 
parameters. A summary of the results of this search -- here showing 
only those runs which possess the optimal $\Psi _{0, \mathrm{Opt}}$ 
value for minimizing $\chi^{2}_{\mathrm{Fit}}$ in each particular 
$z_\mathrm{Init}$ case -- is given in Table~\ref{TableRNLt0SatOptimRuns}.

\placetable{TableRNLt0SatOptimRuns}

We see that the best (minimized) value of $\chi^{2}_{\mathrm{Fit}}$ 
obtainable is in fact somewhat different for different choices of 
the input parameter $z_\mathrm{Init}$; but varying even more dramatically 
with $z_\mathrm{Init}$ is the average deviation from Concordance for 
its $\chi^{2}_{\mathrm{Fit}}$-minimizing run. We have highlighted 
(in bold) five of the cases in Table~\ref{TableRNLt0SatOptimRuns}: 
$\Psi _{\mathrm{MD}}$ with $(z_\mathrm{Init}, \Psi _{0}) = (1.5, 2.8)$ 
and $(z_\mathrm{Init}, \Psi _{0}) = (2, 2.9)$; $\Psi _{\mathrm{Lin}}$ 
with $(z_\mathrm{Init}, \Psi _{0}) = (2, 3.2)$ and 
$(z_\mathrm{Init}, \Psi _{0}) = (3, 3.3)$; and $\Psi _{\mathrm{Sqr}}$ 
with $(z_\mathrm{Init}, \Psi _{0}) = (25, 4.1)$. We will refer to 
these cases as `best runs', chosen not by the global minimization of 
$\chi^{2}_{\mathrm{Fit}}$, but by selecting $z_\mathrm{Init}$ settings 
for which the $\chi^{2}_{\mathrm{Fit}}$-minimizing 
$\Psi _{0, \mathrm{Opt}}$ run produces cosmological parameters with 
very small deviations from Concordance, in addition to having values 
of $\chi^{2}_{\mathrm{Fit}}$ that are acceptably close to the smallest 
ones found for the relevant $\Psi (t)$ function. We will consider 
the astrophysical implications of the particular 
$(z_\mathrm{Init},\Psi _{0, \mathrm{Opt}})$ input values required 
to obtain these `best' runs, later on, below; but first, 
we consider here the principal output results of interest. 

\begin{figure}
\begin{center}
\includegraphics[scale=1.0]{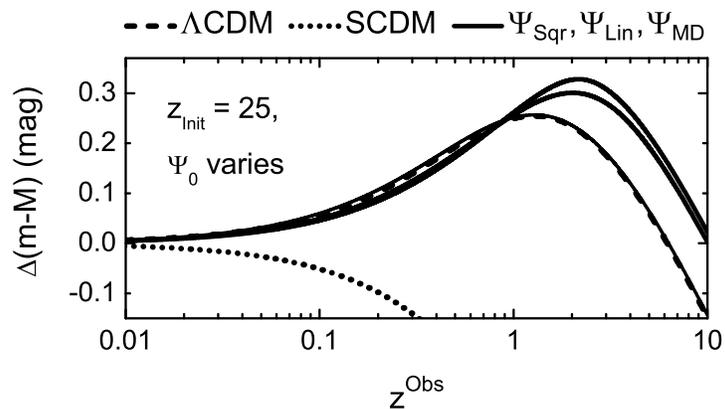}
\end{center}
\caption{Residual Hubble diagrams for three of the `best' runs 
with $\Psi _{0} > 1$ (as described in the text), using 
model input parameters (solid lines increasing from lowest to highest): 
$\Psi _{\mathrm{Sqr}}$ with $(z_\mathrm{Init},\Psi _{0}) = (25,4.1)$ 
(curve almost indistinguishable from $\Lambda$CDM); 
$\Psi _{\mathrm{Lin}}$ with $(z_\mathrm{Init},\Psi _{0}) = (3,3.3)$; 
and $\Psi _{\mathrm{MD}}$ with $(z_\mathrm{Init},\Psi _{0}) = (2,2.9)$. 
Shown along with them for comparison (broken lines) are the flat SCDM 
and (Union1-best-fit, $\Omega _{\Lambda} = 0.713$) Concordance 
$\Lambda{\mathrm{CDM}}$ cosmologies.}
\label{FigCLgt1StrongPlots}
\end{figure}

These five best runs (now re-done with 10000 pixels each, for higher 
precision) are all good fits to the SNIa data, as can be seen from 
their $\chi^{2}_{\mathrm{Fit}}$ values, which are nearly as 
small as that for the best-fit $\Lambda$CDM model (and are even smaller 
than that from $\Lambda$CDM for {\it all} of the $\Psi _{\mathrm{Sqr}}$ 
runs in the table). The high quality of these fits can also be inferred 
from Figure~\ref{FigCLgt1StrongPlots}, where we plot three of these runs 
-- one each for $\Psi _{\mathrm{MD}}$, $\Psi _{\mathrm{Lin}}$, and 
$\Psi _{\mathrm{Sqr}}$ -- against $\Lambda$CDM and the matter-only flat 
SCDM model, showing our runs to be very good at mimicking best-fit 
$\Lambda$CDM (particularly within the redshift region $z \lesssim 1$, 
containing most of the SNIa), thus demonstrating their ability to 
successfully reproduce the cosmic `acceleration' as it is actually observed.

\placefigure{FigCLgt1StrongPlots}

The derived output parameters for these five best $\Psi _{0} > 1$ 
runs are presented in Table~\ref{TableRNLt0SatBestRuns}. Here we see 
how their $\chi^{2}_{\mathrm{Fit}}$ values, nearly as good or (for 
$\Psi _{\mathrm{Sqr}}$) better than that for best-fit $\Lambda$CDM, 
makes them comparable to $\Lambda$CDM in terms of the fit probability, 
$P_{\mathrm{Fit}}$ (noting that $\Lambda$CDM has fewer optimizable 
model parameters, and thus more degrees of freedom, giving it a higher 
$P_{\mathrm{Fit}}$ value for the same $\chi^{2}_{\mathrm{Fit}}$). 
We also see that their present-day `Newtonian' perturbation values, 
$I_{0} \equiv I^{\mathrm{RNL}}(t_{0})$, do not go much higher than 
$I_{0} \sim 0.6$; this places moderate limits upon the effects of 
gravitational nonlinearities -- which are {\it not} explicitly dealt 
with in our formalism -- implying at least some reasonable level 
of accuracy for our models regarding this approximation.

\placetable{TableRNLt0SatBestRuns}

Considering the observational cosmological parameters, there is an 
astrophysically-interesting modification of $z^\mathrm{Obs}$ (relative to 
$z^\mathrm{FRW} = 1$) of $\sim$$14 - 17 \%$; while the best-fit observable 
Hubble constant, $H^\mathrm{Obs}_{0}$, is in all cases extremely close 
(well within $1 \%$) of the $\Lambda$CDM Concordance value of 
$69.96 ~ \mathrm{km} ~ \mathrm{s}^{-1} \mathrm{Mpc}^{-1}$. The cosmic ages 
of these models are all at least equal to the ``Concordance'' value of 
$t^\mathrm{Obs}_{0} = 13.64$ GYr (with the $\Psi _{\mathrm{Sqr}}$ run being 
{\it exactly} equal to it); and since cosmologies with higher (if still 
reasonable) ages would also solve the Cosmic Age Problem just as well, all of 
these runs (no more than $\sim$$0.8$ GYr older) therefore appear to have quite 
acceptable values of $t^\mathrm{Obs}_{0}$. 

The goal of achieving spatial flatness without Dark Energy 
is also accomplished quite successfully, with 
$\Omega^\mathrm{FRW}_\mathrm{M} \sim 0.94 - 1.16$. This result is even 
more impressive given the fact that to produce these numbers, one must 
first normalize the calculations to some pre-specified value of the 
observed matter density, $\Omega^\mathrm{Obs}_\mathrm{M}$; and thus any 
uncertainties in that observed density will translate themselves into 
error bars on the $\Omega^\mathrm{FRW}_\mathrm{M}$ values output from 
the simulations -- and with a factor {\it greater} than unity, since 
$\Delta \Omega^\mathrm{FRW}_\mathrm{M} 
\simeq \Delta \Omega^\mathrm{Obs}_\mathrm{M} 
(\Omega^\mathrm{FRW}_\mathrm{M} / \Omega^\mathrm{Obs}_\mathrm{M}) \sim 
\Delta \Omega^\mathrm{Obs}_\mathrm{M} (1 / \Omega^\mathrm{Obs}_\mathrm{M}) 
\sim (3-4) \times \Delta \Omega^\mathrm{Obs}_\mathrm{M}$. For all of the 
results presented in this paper, we have used 
$\Omega^\mathrm{Obs}_\mathrm{M} \equiv 0.27$. If one assumes that this value 
is more likely to be too high than too low, then models here with results of 
$\Omega^\mathrm{FRW}_\mathrm{M} \gtrsim 1.0$ would be somewhat better than 
those with $\Omega^\mathrm{FRW}_\mathrm{M} \lesssim 1.0$; but in any case, 
all of the models in Table~\ref{TableRNLt0SatBestRuns} are clearly 
successful at effectively achieving flatness. 

In terms of providing a sufficient (apparent) acceleration, these models 
are again successful, with $w^\mathrm{Obs}_{0} \sim (-0.64) - (-0.75)$, 
effectively bracketing the Concordance value of $w^\mathrm{Obs}_{0} = -0.713$. 
We also note that since the precise cosmological parameters obtained from SNIa 
fits are highly model-dependent \citep[e.g.,][]{CattVisserCosmographSNeFits}, 
it is therefore not necessary for alternative-cosmology models to {\it exactly} 
reproduce the $w^\mathrm{Obs}_{0}$ result from Concordance $\Lambda$CDM, but 
merely to generate an `accelerative' value of $w^\mathrm{Obs}_{0}$ strong 
enough to provide a sufficiently good fit to the supernova data, as has been 
done here. 

One final quantity of great importance to be reproduced by an alternative 
concordance is the CMB acoustic scale, $l^\mathrm{Obs}_{\mathrm{A}}$ -- a 
cosmological parameter that weighs the cosmic evolution over vast distances 
in space and time. We see here that for the Concordance value of 
$l^\mathrm{Obs}_{\mathrm{A}} = 285.4$ (as computed using a toy $\Lambda$CDM 
model with no radiation), excellent matches are achieved by 
$\Psi _{\mathrm{MD}}$ with $(z_\mathrm{Init}, \Psi _{0}) = (2, 2.9)$, and 
$\Psi _{\mathrm{Lin}}$ with $(z_\mathrm{Init}, \Psi _{0}) = (3, 3.3)$ (which 
is why we chose these $\Psi _{\mathrm{MD}}$ and $\Psi _{\mathrm{Lin}}$ models 
to show in Figure~\ref{FigCLgt1StrongPlots} over the other highlighted ones 
in Table~\ref{TableRNLt0SatOptimRuns}, despite their slightly worse 
$\chi^{2}_{\mathrm{Fit}}$ values); and sufficiently good matches are obtained 
by the remaining three models in Table~\ref{TableRNLt0SatBestRuns}. Although 
a discrepancy of $\Delta l^\mathrm{Obs}_{\mathrm{A}} \sim 10$ with respect 
to the value expected from $\Lambda$CDM Concordance is certainly something 
that would be noticeable from observations, we will see later on 
(in Section~\ref{SubEarlySatClumping}) that one can do even better than this 
with more refined models (for the $\Psi _{\mathrm{Sqr}}$ case, particularly). 

The last remaining output parameter quoted in 
Table~\ref{TableRNLt0SatBestRuns} is the ``jerk'' or ``jolt'' parameter, 
$j^\mathrm{Obs}_{0}$. As discussed at length in BBI, rather than serving 
as a well-observed cosmological parameter to be matched by theories aiming 
for an alternative concordance, the (still poorly-measured) value of 
$j^\mathrm{Obs}_{0}$ can be viewed as a {\it prediction} of one's theory, 
useful for ultimately distinguishing such theories from Cosmological Constant 
Dark Energy (since flat $\Lambda$CDM is always constrained to $j_{0} = 1$, 
for {\it any} value of $\Omega _{\Lambda} = 1 - \Omega _\mathrm{M}$). We 
recall that the old version of our causal backreaction simulation program, 
without RNL, obtained a set of best-fit models which all possessed very high 
values of the jerk parameter -- specifically, $j_{0} \sim 2.6 - 3.8$ for the 
half-dozen best runs -- providing a clear and distinct way of observationally 
distinguishing our causal backreaction paradigm from a Cosmological Constant, 
or from any other slowly-evolving form of Dark Energy reasonably close to 
$\Lambda$CDM. But here the results in Table~\ref{TableRNLt0SatBestRuns} show 
that when RNL is properly incorporated into our formalism, the situation 
becomes far less clear: the best-fitting $\Psi _{\mathrm{MD}}$ and 
$\Psi _{\mathrm{Lin}}$ models now prefer {\it low} values of the 
jerk parameter ($j^\mathrm{Obs}_{0} \sim 0.5 - 0.6$ and 
$j^\mathrm{Obs}_{0} \sim 0.8 - 0.9$, respectively); and the best-fit 
$\Psi _{\mathrm{Sqr}}$ model, though still predicting a value higher than unity, 
now only goes so high as $j^\mathrm{Obs}_{0} \sim 1.7$. A further analysis 
below in Section~\ref{SubEarlySatClumping} will demonstrate that even this 
result may be subject to substantial change; and thus the search for a reliable 
observational method for `falsifying' our causal backreaction paradigm (that is, 
materially distinguishing it from $\Lambda$CDM) will likely be more difficult 
than what was hoped for in BBI, where the possible impact of recursive 
nonlinearities was still only dealt with in terms of qualitative caveats. 

With this analysis of our `output' results done, we now look at the 
{\it input} parameters which were necessary to produce these good 
cosmological fits. First of all, considering $z_\mathrm{Init}$, we see 
from Table~\ref{TableRNLt0SatOptimRuns} that to obtain good cosmological 
output parameters from the $\Psi _{\mathrm{MD}}$ and $\Psi _{\mathrm{Lin}}$ 
models, we have had to impose an {\it extremely late} value for the time 
representing the `beginning' of structure formation -- specifically, 
$z_\mathrm{Init} \sim 1.5 - 3$. This is a direct result of the effects of 
RNL, which for an already-perturbed universe will slow down the continued 
propagation of old perturbations coming in from great cosmological distances, 
thus rendering the causal backreaction contributions of early-developing 
inhomogeneities to the late-time `acceleration' much weaker than they were 
for the equivalent models simulated in BBI. This is a serious problem for 
the $\Psi _{\mathrm{Lin}}$ and (even more so) for the $\Psi _{\mathrm{MD}}$ 
functions, which are especially designed to provide more early clustering, 
at the expense of late-time clustering. We must ask, in fact, whether input 
values like these for $z_\mathrm{Init}$ even make astrophysical sense, 
given the fact that $z^{\mathrm{Obs}} \sim 3$ actually marks the epoch 
during which the large-scale collapse of material and the resulting 
galactic feedback (from star formation, etc.) begins acting strongly 
to shock-heat cosmic baryons, sending a significant portion of the baryons 
into the superheated IGM \citep[e.g.,][]{CenOstrikerShockedOb}, 
thus helping to {\it slow down} further structure formation. 

This problem goes away, however, for the $\Psi _{\mathrm{Sqr}}$ runs, which 
have enough late-time clumping to remove the need for choosing a late (i.e., 
small) value of $z_\mathrm{Init}$ to get good output cosmological parameters. 
The runs only get better, in fact, as one goes to higher $z_\mathrm{Init}$, 
and we stop at $z_\mathrm{Init} = 25$ not because optimization tells us to 
(though increasingly early initialization times eventually brings diminishing 
returns), but because it is not physically reasonable to set the start of 
clustering too long before the onset of reionization. 

Besides this success of providing an alternative concordance with 
reasonable $z_\mathrm{Init}$ values, some reflection shows that the 
$\Psi _{\mathrm{Sqr}}$ runs are the most sensible in every regard. As we 
now consider the $\Psi(t)$ function to represent the feedback effects of 
completed clustering on several different, dynamically-defined scales 
in hierarchical structure formation -- rather than as a simple percentage 
of `clumped' versus `unclumped' matter -- the causal backreaction process 
becomes more naturally focused upon structure formation in the highly 
nonlinear regime of density perturbations. And as we recall from above, 
density fluctuations in the nonlinear regime evolve as 
$\delta \rho / \rho \propto t^{m}$ with $m \gtrsim 2$ for a 
matter-dominated universe, meaning that $\Psi(t)$ functions evolving as 
$\Psi _{\mathrm{Sqr}} \propto t^{2}$ (or functions growing with even 
larger powers of $t$) are therefore the most relevant models for 
causal backreaction, rather than those with smaller powers of $t$ like 
the $\Psi _{\mathrm{Lin}}$ or $\Psi _{\mathrm{MD}}$ models (the latter 
type evolving as $\delta \rho / \rho$ in the {\it linear} regime), 
which had been the dominant ones for generating backreaction 
in the old simulations without RNL. 

Considering the above results once more in this light, we see 
that $\Psi _{\mathrm{Sqr}}$ has indeed performed better than 
$\Psi _{\mathrm{Lin}}$ or $\Psi _{\mathrm{MD}}$, for example 
looking much more like best-fit $\Lambda$CDM in 
Figure~\ref{FigCLgt1StrongPlots}, thus providing better 
SNIa fits with significantly lower $\chi^{2}_{\mathrm{Fit}}$ values 
(recall Tables~\ref{TableRNLt0SatOptimRuns},\ref{TableRNLt0SatBestRuns}). 
Furthermore, the output cosmological parameters from the 
$\Psi_{0}$-optimized $\Psi _{\mathrm{Sqr}}$ runs (with sufficiently 
large $z_\mathrm{Init}$ values) were all extremely good, with just 
the possible exceptions of a slightly low total density 
(e.g., $\Omega^\mathrm{FRW}_\mathrm{M} \sim 0.94$ for 
$z_\mathrm{Init} = 25$), and a slightly errant CMB angular scale 
($\Delta l^\mathrm{Obs}_{\mathrm{A}} \sim 9$); but even those 
discrepancies are quite small given the very simplified nature of 
these clumping evolution models. (And those discrepancies can be 
made even smaller with some further development of the models, 
as will be seen shortly, in Section~\ref{SubEarlySatClumping}). 

The only real problem with these $\Psi _{\mathrm{Sqr}}$ models, and 
quite an obvious one, is the very high degree of clumping needed in 
order to generate these results -- that is, 
$\Psi _{0, \mathrm{Opt}} \simeq 4.1$. This is right around the extreme 
upper edge of the `reasonable' range of values for $\Psi _{0}$, which 
we previously argued should likely lie within $\Psi _{0} \sim 2 - 4$ 
(and more conservatively, within $\Psi _{0} \sim 2 - 3$). Such 
$\Psi _{0, \mathrm{Opt}}$ values for $\Psi _{\mathrm{Sqr}}$ (which are 
greater than those for $\Psi _{\mathrm{Lin}}$ or $\Psi _{\mathrm{MD}}$) 
are large enough as to potentially strain the credibility of causal 
backreaction as a cosmological solution. As we will see below, however, 
this concern can be alleviated simply by adding in one important and 
physically reasonable model parameter as part of these clumping 
evolution functions: specifically, a parameter representing information 
about the possible {\it end} of clumping, as could be triggered in 
the real universe by galactic feedback.

\subsection{\label{SubEarlySatClumping}Early-Saturation Clustering Models: 
Parameter Flexibility and Paradigm Falsifiability} 

As our clumping evolution models have so far been defined, each takes on its 
`ultimate' value, $\Psi _{0}$, at $t_{0}$; and this may make physical sense 
in a universe where the effective clumpiness is still growing, so that the 
value $\Psi (t_{0})$ is not particularly special, but just represents a 
`snapshot' of $\Psi (t)$ taken at our particular observation time. But if 
there is instead some halt to the growth of clumping -- due either to the 
limiting value of $\Psi (t) \rightarrow 1$ for our pre-RNL simulations, or 
due to various feedback effects in these new models (which no longer possess 
a pre-determined upper bound on $\Psi (t)$) -- then from a Copernican view, 
there is nothing necessarily special about the current cosmic epoch, 
and so there is no reason to assume that the asymptotic value of $\Psi (t)$ 
(if there is one) should occur specifically at the present time. 

Having already re-evaluated our formalism by interpreting the implementation 
of $\Psi _{0} > 1$ as representing multiple levels of structure on different 
cosmic scales, it has become possible for $\Psi (t)$ to continue increasing 
without any hard theoretical limit for $t > t_{0}$, perhaps considerably 
beyond the effective degree of clumping that exists today. On the other hand, 
if there {\it does} happen to be some naturally-occurring, effective limit 
of clumping, then there is no reason to believe that the processes in charge 
of cosmic clustering would wait until $t \rightarrow t_{0}$ to reach it. 

As noted above, \citet{CenOstrikerShockedOb} show that the nonlinear 
collapse of material and feedback from star formation have likely acted 
to shock-heat cosmic baryons to millions of degrees, thus inhibiting 
clumping -- the baryonic part of it, at the very least -- by keeping 
and/or sending a significant portion of the baryons into the superheated 
IGM; their Figure 1 shows simulation results depicting the rapid increase 
of very hot baryons in the IGM beginning strongly as of 
$z^{\mathrm{Obs}} \sim 3$. 

Additionally, the backreaction effect itself may be helping to put the 
brakes on structure formation at late times; particularly given the fact 
that it is generated most strongly within the vicinity of the virializing 
masses themselves, in high-density regions. In that case, signs of a 
cosmologically recent slow-down in the growth of clustering 
\citep[e.g.,][]{VikhDEevo} --  findings which are viewed as strong evidence 
in favor of Dark Energy -- may instead be indicating the effects of causal 
backreaction. (Conceivably, feedback from backreaction may also be a 
contributor to other poorly understood cosmological effects such as galaxy 
downsizing \citep{CowieDownsizing}, the cuspy CDM halo problem, and the 
possible dearth of dwarf satellite galaxies \citep[e.g.,][]{PrimackCuspy}.) 

What all of this implies is that our clumping evolution functions, 
as they have been defined in Equations~\ref{EqnClumpModels}a-c, would 
likely not remain meaningful all the way to $t \rightarrow t_{0}$; 
and thus it is useful to consider how the functional form of our models, 
$\Psi (t)$, might be altered to account in some manner for this effect, 
to represent the possible late-time softening of the growth of clustering 
due to astrophysical and/or backreaction-related feedback processes. 

The most straightforward way to model this behavior is by implementing 
an early ``saturation time'', $t_\mathrm{Sat} < t_{0}$ (with 
$z_\mathrm{Sat} = [(t_{0}/t_\mathrm{Sat})^{2/3} - 1]$), representing 
the time {\it before} the present epoch at which point the growth of 
clustering became saturated to a final, fixed value, 
$\Psi (t_\mathrm{Sat}) \equiv \Psi _{0}$. This new clumping evolution 
function then fixes $\Psi (t)$ to remain equal to $\Psi _{0}$ all the 
way from $t_\mathrm{Sat}$ to $t_{0}$. Recalling from 
Section~\ref{SubSuperUnityClumpStrong} that the $\Psi _{\mathrm{MD}}$ 
and $\Psi _{\mathrm{Lin}}$ models only obtain their best fits -- and 
their most `concordant' cosmological parameters -- for very late 
$t_\mathrm{Init}$ values (i.e., $z_\mathrm{Init} \sim 1.5 - 3$), it 
then makes little sense to outfit those models with an even later 
$t_\mathrm{Sat} > t_\mathrm{Init}$; we therefore make this modification 
only for the $\Psi _{\mathrm{Sqr}}$ model, which was optimized for 
concordance with much earlier $t_\mathrm{Init}$ values. The resulting, 
retrofitted $\Psi _{\mathrm{Sqr}}$ function for early saturation 
is thus defined as: 
\begin{equation}
\Psi _{\mathrm{Sat}} (t) \equiv \left\{
\begin{array}{rl}
0 \phantom{ABC1} & \textrm{for } ~ t \le t_\mathrm{Init} ~,\\ 
\Psi _{0} ~ (\frac{t - t_\mathrm{Init}}{t_\mathrm{Sat} - t_\mathrm{Init}})^{2} 
& \textrm{for } ~ t_\mathrm{Init} < t < t_\mathrm{Sat} ~,\\ 
\Psi _{0} \phantom{ABC} & \textrm{for } ~ t \ge t_\mathrm{Sat} ~.
\end{array} \right.
\label{EqnPsiSatDefn}
\end{equation}
This is obviously a very elementary way of accounting for how astrophysical 
feedback acts to slow down the growth of clustering (with the resultant 
lessening of its causal backreaction); but even such an oversimplified 
model will provide us with valuable cosmological lessons. 

As a first run to explore the implications of early saturation, we consider 
again our earlier, cosmologically successful run with $\Psi _{\mathrm{Sqr}}$ 
(functionally identical to $\Psi _{\mathrm{Sat}}$ with $z_\mathrm{Sat} = 0$), 
which had used $z_\mathrm{Init} = 25$ and its corresponding 
$\Psi _{0, \mathrm{Opt}} = 4.1$; these results are now compared to those 
from a new $\Psi _{\mathrm{Sat}}$ run still using $z_\mathrm{Init} = 25$ 
and (the now non-optimal) $\Psi _{0} = 4.1$, but with $z_\mathrm{Sat} = 3$ 
now imposed -- a conservatively early value suggested by the onset of 
substantial hot baryon injection into the IGM. 

\begin{figure}
\begin{center}
\includegraphics[scale=1.0]{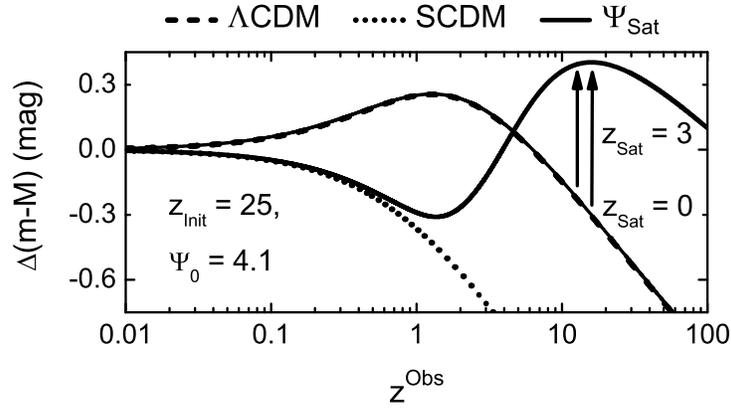}
\end{center}
\caption{Residual Hubble diagrams for the early-saturation 
causal backreaction model $\Psi _{\mathrm{Sat}} (t)$ with 
$(z_\mathrm{Init},\Psi _{0}) = (25,4.1)$, where the 
solid line peaking at $z^\mathrm{Obs} \sim 1 - 2$ 
(the curve almost indistinguishable from $\Lambda$CDM)
represents a simulated cosmology with RNL but restricted 
to $z_\mathrm{Sat} = 0$ (thus rendering it equivalent 
to $\Psi _{\mathrm{Sqr}}$); and where the solid line 
peaking at $z^\mathrm{Obs} \sim 10 - 20$ represents the 
shift to $z_\mathrm{Sat} = 3$. Shown along with them for 
comparison (broken lines) are the flat SCDM and 
(Union1-best-fit, $\Omega _{\Lambda} = 0.713$) 
Concordance $\Lambda{\mathrm{CDM}}$ cosmologies.}
\label{FigEarlyVsLateSat}
\end{figure}

The resulting residual Hubble diagrams for the two cases are compared in 
Figure~\ref{FigEarlyVsLateSat}, where the powerful impact of moving from 
$z_\mathrm{Sat} = 0$ to $z_\mathrm{Sat} = 3$ upon the time-dependence of 
the observed backreaction can clearly be seen. Not unexpected is the 
strongly-amplified apparent acceleration effect at early times 
($z^\mathrm{Obs} \sim 10$), due to the greatly compressed timescale over 
which clustering grows from $\Psi (t) = 0$ to $\Psi (t) = \Psi _{0} = 4.1$. 
But what may be surprising is how dramatically the effects of these 
early-developing inhomogeneities become virtually irrelevant at late times, 
with the effective `acceleration' almost completely fading out by 
$z^\mathrm{Obs} \sim 1 - 2$, and the residual Hubble diagram relapsing 
to a nearly perfect SCDM (i.e., decelerating) cosmology not long after 
$z^\mathrm{Obs} \sim 1$.

\placefigure{FigEarlyVsLateSat}

This is in sharp contrast to what would be expected from the 
results explored in BBI, where early-developing clumping provided the 
strongest ongoing causal backreaction effects due to the ever-growing 
``inhomogeneity horizons'' of observers. The fact that this is no 
longer the case is clearly due to RNL, which damps such ongoing 
backreaction effects by slowing down the causal propagation of 
inhomogeneity information (as well as by diluting the perturbation 
effects of already-seen inhomogeneities via the extra volumetric 
expansion), thus greatly restricting the cosmological horizon out 
from which developing perturbations can affect the observer. As can 
be inferred from 
Equations~\ref{EqnRofTIntegration},\ref{EqnAlphaRNLintegration2} above, 
$d r / dt \propto \sqrt{1 - I^{\mathrm{RNL}}(t)}$ for inhomogeneity 
information propagating towards an observer at null-ray speed; 
and so models with very strong early clumping -- particularly those 
with a very early $z_\mathrm{Sat}$ -- drive up the value of 
$I^{\mathrm{RNL}}(t)$ so close to unity, so early on, that the 
propagation of inhomogeneity information is practically frozen to a 
halt at later times. The cessation of new inhomogeneity information 
reaching the observer leads to a corresponding freeze in the continued 
evolution of $I^{\mathrm{RNL}}(t)$; and as is obvious from our metric 
given by Equation~\ref{EqnFinalBHpertMetric}, a constant 
$I^{\mathrm{RNL}}$ value can simply be transformed away via redefinitions 
of $t$ and $\vec{r}$, thus making a cosmology with static 
$I(t) = I^{\mathrm{RNL}}$ look exactly like a decelerating, 
matter-dominated SCDM universe. (Interestingly, a considerably stronger 
late-time acceleration effect can be generated by using a {\it smaller} 
value of $\Psi _{0}$ for these very early $z_\mathrm{Sat}$ runs, which 
lessens this `freezing' effect. The best possible SNIa fit for this 
$z_\mathrm{Sat} = 3$ case is therefore achieved with the relatively 
low value of $\Psi _{0, \mathrm{Opt}} = 1.2$; though even that remains 
a very poor fit in absolute terms.)

The importance of such behavior is that it limits the degree to which 
causal backreaction with RNL can be `self-powered' via the ever-expanding 
reach of observational horizons, in defiance of a static final $\Psi _{0}$ 
value; a result which increases its dependence upon being `driven' by a 
continually-growing $\Psi (t)$ function. As Figure~\ref{FigEarlyVsLateSat} 
has shown, an early saturation of clumping can lead to a virtually complete 
shutdown of apparent acceleration not much later. (Though noting again the 
caveat that nonlinear gravitational effects are not accounted for in any of 
our models -- and that they may indeed be important here, since all of the 
very-early saturation models with $z_\mathrm{Sat} = 3$ which we tested do 
have excessively large $I_{0}$ values, all significantly exceeding $0.7$, 
with some nearly approaching unity.) These considerations will have 
important implications for the eventual fate of the universe, as will 
be discussed below in Section~\ref{SubNewFuture}; but first, we consider 
their ramifications regarding the {\it past} cosmic `acceleration' that 
has already been observed via the supernova data sets. 

Since a value of $z_\mathrm{Sat} = 3$ effectively terminates the impact of 
causal backreaction far too prematurely for it to account for the apparent 
acceleration seen in the SNIa data from around $z^\mathrm{Obs} \sim 1$ to now, 
we consider runs with $z_\mathrm{Sat}$ values that are similar to or smaller 
(i.e., later) than this epoch of acceleration. Specifically, we choose values 
of $z_\mathrm{Sat} \equiv z^{\mathrm{FRW}}_\mathrm{Sat} = (1.0,0.5,0.25)$ 
for study -- i.e., $t^{\mathrm{FRW}}_\mathrm{Sat} \simeq (0.35,0.5,0.7)$ -- 
representing times by which the mass fraction of cosmic baryons possessing 
temperatures of $T > 10^{5}$K has increased to reach 
$\sim$$0.4 - 0.5$ \citep[][Figure 1b]{CenOstrikerShockedOb}. 
These values of $z_\mathrm{Sat}$ are also well-placed to further illuminate 
the results of \citet{VikhDEevo}, which demonstrated the significant 
alteration in clustering behavior (due either to Dark Energy or to 
causal backreaction) between one cluster sample with $z > 0.35$ (and 
particularly its most distant subsample with $z > 0.55$), versus their 
more nearby cluster sample at $z < 0.25$. 

Re-optimizing $\Psi _{0}$ for each of these chosen $z_\mathrm{Sat}$ 
values (with $z_\mathrm{Init} = 25$ fixed for all of these runs), 
the resulting `best runs' with early saturation are those with the 
parameters: $(z_\mathrm{Sat},\Psi _{0}) = (0.25,2.6)$, 
$(z_\mathrm{Sat},\Psi _{0}) = (0.5,2.3)$, 
and $(z_\mathrm{Sat},\Psi _{0}) = (1.0,2.2)$; 
we compare this to the best-fitting $z_\mathrm{Sat} = 0$ run, 
which had $\Psi _{0, \mathrm{Opt}} = 4.1$. 

\begin{figure}
\begin{center}
\includegraphics[scale=1.0]{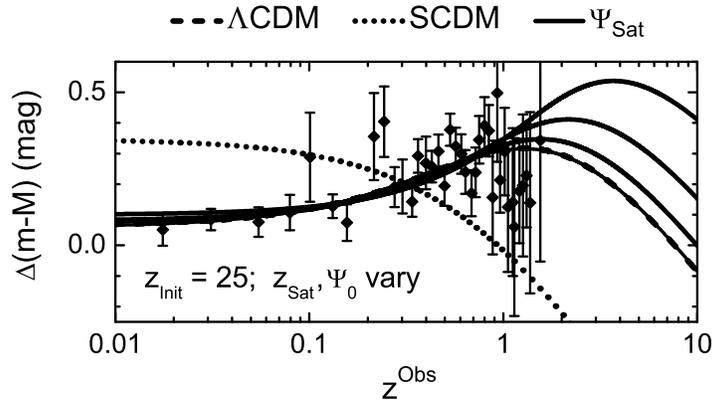}
\end{center}
\caption{
Residual Hubble diagrams for early-saturation 
$\Psi _{\mathrm{Sat}} (t)$ models with fixed 
$z_\mathrm{Init} = 25$ and varying $z_\mathrm{Sat}$, 
with $\Psi _{0}$ re-optimized for each chosen 
$z_\mathrm{Sat}$ value. From lowest to highest, 
the solid lines represent the runs with model input 
parameters: $(z_\mathrm{Sat},\Psi _{0}) = (0,4.1)$ 
(curve almost indistinguishable from $\Lambda$CDM); 
$(z_\mathrm{Sat},\Psi _{0}) = (0.25,2.6)$; 
$(z_\mathrm{Sat},\Psi _{0}) = (0.5,2.3)$; 
and $(z_\mathrm{Sat},\Psi _{0}) = (1.0,2.2)$. 
Also plotted here (broken lines) are the Union1-best-fit 
flat SCDM and Concordance $\Lambda{\mathrm{CDM}}$ 
($\Omega _{\Lambda} = 0.713$) cosmologies. 
Shown along with these curves are the 
binned and averaged SCP Union1 SNIa data 
(bin size $\Delta \mathrm{Log}_{10} [ 1 + z ] = 0.01$). 
Each theoretical curve is displaced vertically, 
relative to the SNIa data, to depict its individualized 
$H^\mathrm{Obs}_{0}$-optimization.}
\label{FigVaryingEarlySatZval}
\end{figure}

Residual Hubble curves for these four runs are plotted in 
Figure~\ref{FigVaryingEarlySatZval}, where we see that while these 
early-saturation runs are no longer effectively indistinguishable 
from best-fit $\Lambda$CDM (as was the $z_\mathrm{Sat} = 0$ run), 
they are nevertheless very close to it -- particularly within the 
redshift range containing most of the SNIa -- thus providing 
quite good fits to these data.

\placefigure{FigVaryingEarlySatZval}

The complete fit quality results and cosmological output parameters for 
these runs are presented in Table~\ref{TableRNLearlySatRuns}. First, 
we note that there are limits to the largest value of $z_\mathrm{Sat}$ 
which may practically be used, since we see that as $z_\mathrm{Sat}$ gets 
as high as $1.0$, the fit probability $P_{\mathrm{Fit}}$ gets progressively 
worse (i.e., $\chi^{2}_{\mathrm{Fit}}$ begins to grow unacceptably large), 
the cosmological parameters become increasingly `discordant' (particularly 
$\Omega^\mathrm{FRW}_\mathrm{M}$ and $l^\mathrm{Obs}_{\mathrm{A}}$), and there 
are even concerns that the formalism itself begins to break down due to 
gravitationally nonlinear effects, given the disturbingly large value of $I_{0}$.

\placetable{TableRNLearlySatRuns}

If we restrict ourselves to $z_\mathrm{Sat} \lesssim 0.5$, however, 
then at a cost of only a small increase in $\chi^{2}_{\mathrm{Fit}}$, the 
situation gets considerably better with the use of nonzero $z_\mathrm{Sat}$ 
in several crucial ways. Considering the $z_\mathrm{Sat} = 0.25$ case 
in particular, moving away from $z_\mathrm{Sat} = 0$ has improved the 
match to the CMB data and the verification of spatial flatness (better 
$l^\mathrm{Obs}_{\mathrm{A}}$ and $\Omega^\mathrm{FRW}_\mathrm{M}$ values, 
respectively), while remaining essentially as good in terms of the other 
cosmological parameters, with just a tiny degradation in the fit probability 
(by $\Delta P_{\mathrm{Fit}} = -0.025$). But most importantly, we see 
that all of this been achieved with a {\it much} lower $\Psi _{0}$ value 
-- Table~\ref{TableRNLearlySatRuns} showing how $\Psi _{0, \mathrm{Opt}}$ 
decreases with increasing $z_\mathrm{Sat}$, in general -- dropping all 
the way from $\Psi _{0, \mathrm{Opt}} = 4.1$ for $z_\mathrm{Sat} = 0$, 
to $\Psi _{0, \mathrm{Opt}} = 2.6$ for $z_\mathrm{Sat} = 0.25$. This is 
now well within the $\Psi _{0} \sim 2-3$ range of `reasonable' values 
for hierarchical clustering, as was specified earlier in 
Section~\ref{SubSuperUnityClumpStrong}. 

(In addition, we have also conducted fits to a more recent supernova 
data set, the SCP Union2 SNIa compilation \citep{AmanRubinSCPunion2}; and 
when we minimize $\chi^{2}_{\mathrm{Fit}}$ for the $z_\mathrm{Sat} = 0.25$ 
case with respect to the Union2 data, rather than the Union1 data, we still 
get similar results: an excellent fit with $\chi^{2}_{\mathrm{Fit}} = 543.6$ 
(compared to $\chi^{2}_{\mathrm{Fit}} = 542.7$ for $\Lambda$CDM)
and $P_{\mathrm{Fit}} \simeq 0.6$; only a slightly higher clumping strength 
value of $\Psi _{0, \mathrm{Opt}} = 2.8$; and similar cosmological parameters 
that are also quite acceptable. Only the unperturbed matter density is 
slightly high with $\Omega^\mathrm{FRW}_\mathrm{M} = 1.13$; 
but as discussed above, this may be due to the adopted value of 
$\Omega^\mathrm{Obs}_\mathrm{M} \equiv 0.27$ being slightly too large.)

In consequence, it is justified to say that we have in every sense 
re-obtained a successful alternative concordance with causal backreaction 
in the presence of recursive nonlinearities, having generated the proper 
amount (and temporal behavior) of apparent acceleration -- as evidenced by 
a fit to the SNIa data that is very nearly as good as that achievable with 
best-fit $\Lambda$CDM (even {\it without} performing a rigorous 
$\chi^{2}_{\mathrm{Fit}}$-minimization search over our model parameter space) 
-- as well as having produced output cosmological parameters that are more 
than acceptably consistent with a variety of complementary cosmic 
measurements. The price to be paid to achieve this goal, due to the 
incorporation of recursive nonlinearities, is the necessity of permitting 
$\Psi _{0}$ values in excess of unity -- which though going against the 
original interpretation of this (heuristic) model parameter, as defined 
in BBI, does seem justifiable in a realistic cosmology exhibiting 
hierarchical structure formation on a variety of scales. On the 
other hand, a new benefit of these results is the shifted focus from 
$\Psi _{\mathrm{MD}}$ models to $\Psi _{\mathrm{Sqr}}$ models, equivalent 
to a shift in emphasis from linearized matter density perturbations to 
nonlinear density perturbations; a new emphasis which in fact makes much 
more sense for a causal backreaction paradigm depending upon vorticity- 
and velocity-dispersion-generated virialization as the fundamental origin 
of structure-induced perturbations to the observed cosmological metric. 

\begin{figure}
\begin{center}
\includegraphics[scale=1.0]{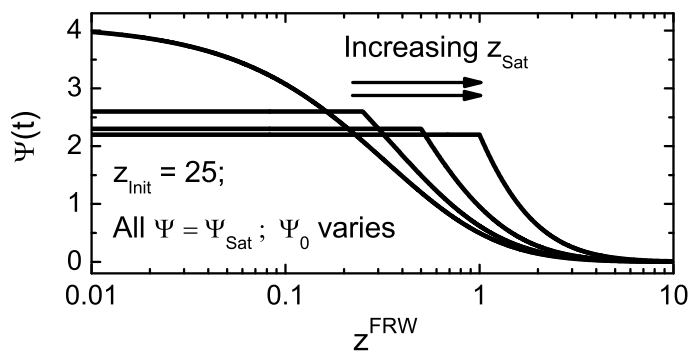}
\end{center}
\caption{The clumping evolution functions are compared for several 
different early-saturation models, all with $z_\mathrm{Init} = 25$, 
plotted versus 
$z^\mathrm{FRW} \equiv [(t^\mathrm{FRW}_{0}/t^\mathrm{FRW})^{2/3} - 1]$. 
In order of the positions of the functions' `corners' from left to right 
(increasing $z_\mathrm{Sat}$), the lines depict $\Psi _{\mathrm{Sat}}$ 
functions with the parameters: $(z_\mathrm{Sat},\Psi _{0}) = (0,4.1)$; 
$(z_\mathrm{Sat},\Psi _{0}) = (0.25,2.6)$; 
$(z_\mathrm{Sat},\Psi _{0}) = (0.5,2.3)$; 
and $(z_\mathrm{Sat},\Psi _{0}) = (1.0,2.2)$.}
\label{FigPsiTplotsVsZsat}
\end{figure}

One last (yet very important) consideration still remains regarding 
early saturation, though, relating to its impact upon the very late-time 
behavior of the cosmic evolution. One way to approach this, is to consider 
{\it why} the optimized value of $\Psi _{0}$ drops so precipitously with 
increasing $z_\mathrm{Sat}$, so conveniently solving our problem by reducing 
$\Psi _{0, \mathrm{Opt}}$ to believable values. To understand why this 
happens, consider the clumping evolution functions, themselves, for the 
early saturation models that we have studied; plots of these $\Psi (t)$ 
functions are shown in Figure~\ref{FigPsiTplotsVsZsat}.

\placefigure{FigPsiTplotsVsZsat}

From these plotted curves, we see that despite the vast differences in 
$\Psi _{0, \mathrm{Opt}}$ for the $z_\mathrm{Sat} = 0$ case versus those 
with $z_\mathrm{Sat} \ne 0$, as long as one does not use too large a value 
of $z_\mathrm{Sat}$ (sticking to, say, $z_\mathrm{Sat} \lesssim 0.5$), 
then the degree of clumping $\Psi (t)$ at {\it mid-range} values of $z$ 
(e.g., $z^\mathrm{FRW} \sim 0.2 - 0.5$) actually remains fairly similar 
from run to run. The distinctively huge increase from $\Psi (t) \sim 2.5$ 
to $\Psi (t) \sim 4$ for the $z_\mathrm{Sat} = 0$ case does not 
actually happen until very late times, $z^\mathrm{FRW} \lesssim 0.2$. 
But why should completely different behaviors of $\Psi (t)$ at late times 
have such a small effect upon the fits to the SNIa data? The answer is 
twofold: first, due to a lack of sensitivity of the data to such changes; 
and second, due to the nature of causal backreaction, itself.

For the first issue, consider that most of the SNIa included in these 
supernova compilations for measuring the cosmic acceleration are located 
at fairly high redshift; it is therefore unsurprising that the evolution 
of $\Psi (t)$ {\it after} the epoch of these supernovae should have little 
effect upon the acceleration measured by those high-$z$ SNIa, regardless of 
whether $\Psi (t)$ experiences a continued increase or a saturation. But, 
even if lower-$z$ SNIa were also included in the study, and could be trusted 
to accurately map out the Hubble flow despite their local peculiar motions, 
there should still be little effect other than a late-time offset, perhaps 
registered as a small change in $H^\mathrm{Obs}_{0}$. As noted by 
\citet{LinderWneg1Mirage}, simple cosmic evolution functions that are 
relatively insensitive to time variations tend to measure an averaged 
cosmological equation of state around a `pivot' redshift of about 
$z^\mathrm{Obs} \approx 0.4$. And so it is understandable why the detailed 
behavior of $\Psi (t)$ after this crucial epoch should end up having little 
effect upon the precise amount of apparent acceleration measured. 

For the second issue, we recall from Equation~\ref{EqnIintegrandPrelim} 
that the amount of causal backreaction due to a spherical shell at 
coordinate radius $r$ will be proportional to 
$d M / r \propto r^{2}/r \propto r$, and so very late clumping -- 
which corresponds to relatively small look-back times $t_{\mathrm{ret}}$, 
and thus small distances $r$ -- will simply not involve a sufficiently 
large amount of inhomogeneous mass to generate significant causal 
backreaction. Thus the detailed dynamics of inhomogeneities at higher 
$z$ (and larger $r$) -- assuming distances not so far away as to get 
damped by RNL, or having look-back times so far back relative to 
$t_\mathrm{Init}$ as to have very small clustering, 
$\Psi (t_{\mathrm{ret}})$ -- will be more important in determining 
the detailed effects of causal backreaction than would the very late 
($z \rightarrow 0$) behavior of clumping. 

The overall result is that there is a parameter degeneracy for causal 
backreaction models, where one can reduce $\Psi _{0}$ in tandem with 
an increase in $z_\mathrm{Sat}$, without much change in the quality of 
the fit to the actual SNIa data. This degeneracy has a beneficial aspect, 
in that it has allowed us to successfully create an alternative concordance 
with causal backreaction -- and to do so in a flat, matter-only universe 
without any form of Dark Energy -- while using astrophysically realistic 
values of $\Psi _{0}$. But it has a negative aspect, as well, in that it 
lessens our ability to predict precise ranges for the late-time cosmological 
parameters that would be output by the various (successfully concordant) 
models in this formalism. Measurable cosmological parameters such as 
$H^\mathrm{Obs}_{0}$, $w^\mathrm{Obs}_{0}$, and $j^\mathrm{Obs}_{0}$ 
are defined -- in theory, at least -- in terms of the 
$z^\mathrm{Obs} \rightarrow 0$ behavior of Taylor expansions of the 
luminosity distance function, $d_\mathrm{L}(z^\mathrm{Obs})$. Hence, while 
a variety of $\Psi (t)$ functions may all provide good fits to the key 
SNIa located at mid-to-high redshifts, these functions may all have 
quite different $t \rightarrow 0$ behaviors, and thus very significant 
differences in their cosmological output parameters. 

In particular, consider again the first three of the four runs in 
Table~\ref{TableRNLearlySatRuns}. As one goes from $z_\mathrm{Sat} = 0$ 
to $z_\mathrm{Sat} = 0.5$, we see that $H^\mathrm{Obs}_{0}$ decreases 
below the $\Lambda$CDM Concordance value of $H^\mathrm{Obs}_{0} \simeq 70$; 
but yet, the difference is small enough to not be a serious concern. 
For $w^\mathrm{Obs}_{0}$, obtained from one more differentiation of 
$d_\mathrm{L}$, the difference is larger, going from 
$w^\mathrm{Obs}_{0} \sim -0.75$ to $w^\mathrm{Obs}_{0} \sim -0.59$; 
and yet, the precise numerical value of the cosmic `equation of state' 
$w^\mathrm{Obs}_{0}$ is of no great concern to us here -- since it is 
not our task to pin down the physics of some hypothesized form of 
Dark Energy, but merely to reproduce the cosmological observations 
-- so long as a good fit with a sufficient amount of apparent 
acceleration can be produced, as has indeed been accomplished. 

The biggest change, however, occurs for $j^\mathrm{Obs}_{0}$ -- obtained 
via yet another differentiation of $d_\mathrm{L}$ -- which drops all the 
way from $j^\mathrm{Obs}_{0} \sim 1.7$ for $z_\mathrm{Sat} = 0$, down 
through the $\Lambda$CDM value of $j_{0} \equiv 1$ and far past it, even 
going negative for $z_\mathrm{Sat}$ values that still provide decent 
SNIa fits. 

A small part of this change may be due to the highly simplified nature 
of the $\Psi _{\mathrm{Sat}}$ functions that we use; in particular, since 
we obtain a Hubble curve through two integrations of $\Psi (t)$ -- that 
is, $\Psi (t) \rightarrow I(t)$, and then $I(t) \rightarrow d_\mathrm{L}$ 
(cf. Equations~\ref{EqnItotIntegration}-\ref{EqnDlumDefn}) -- the 
$j^\mathrm{Obs}_{0}$ parameter (obtained from three differentiations of 
$d_\mathrm{L}$) essentially contains one differentiation of $\Psi (t)$. 
As is obvious from Figure~\ref{FigPsiTplotsVsZsat}, our simple 
$\Psi _{\mathrm{Sat}}$ functions are not differentiable at $z_\mathrm{Sat}$, 
meaning that $q^\mathrm{Obs} (z)$ (and thus $w^\mathrm{Obs} (z)$) will 
also be non-differentiable there, and so $j^\mathrm{Obs} (z)$ will pick 
up an actual discontinuity there. Still, the discontinuity is not large 
enough to account for the huge overall differences in $j^\mathrm{Obs}_{0}$ 
for the different $z_\mathrm{Sat}$ values; and most of the effect of 
the jump in $j^\mathrm{Obs} (z)$ at $z_\mathrm{Sat}$ is almost 
certainly real, due to the real change in the cosmic evolution at 
that point, and would likely still take place to a very similar degree 
(just more spread out in time) for some more complicated 
$\Psi _{\mathrm{Sat}}$ functions designed to apply smoothing at 
$z_\mathrm{Sat}$. In any case, the set of runs in 
Table~\ref{TableRNLearlySatRuns} does fairly clearly establish 
the trend that increasing $z_\mathrm{Sat}$ away from zero leads 
to a steady decrease in the final value of $j^\mathrm{Obs}_{0}$. 

What we can conclude with some certainty, unfortunately, is that 
we have lost any reasonable degree of predictability for this 
$j^\mathrm{Obs}_{0}$ parameter in the causal backreaction paradigm, 
due the weak dependence of the backreaction upon the very-late-time 
behavior of $\Psi (t)$, and the resulting degeneracy in 
$(z_\mathrm{Sat},\Psi _{0})$. This is significant, 
because of the different possible ways suggested in BBI to 
distinguish between causal backreaction and Cosmological Constant 
$\Lambda$CDM (or any similar form of Dark Energy), in order to provide 
our paradigm with a falsifiable test, the clearest signature by 
far was the search for $j^\mathrm{Obs}_{0} \gg 1$. For all intents and 
purposes, the reliability of this signature now appears to be gone, 
and some more intricate means will obviously be needed to distinguish 
causal backreaction from even the simplest, $\Lambda$-like version of 
Dark Energy (though $j^\mathrm{Obs}_{0} \ne 1$ would still rule out 
$\Lambda$CDM in favor of {\it some} alternative model). Thus while 
the use of early saturation has greatly enhanced the prospects for 
achieving an alternative cosmic concordance with realistic values 
of the clumping parameter $\Psi _{0}$, the incorporation of this 
additional degree of freedom in the models has made it significantly 
more challenging to definitively replace the paradigm of Dark Energy 
with that of causal backreaction, without resorting to 
aesthetic arguments of a subjective nature.

\subsection{\label{SubNewFuture}The New Cosmic Future: How Recursive 
                   Nonlinearities Affect the Possibility of 
                   Long-Term Acceleration from Causal Backreaction} 

If appropriate observational tests should eventually be able to 
demonstrate causal backreaction as superior to Dark Energy as the 
driver of the cosmic evolution, then one final question of great 
importance would of course be: What is the ultimate fate of 
the universe? 

This topic was discussed in detail in BBI, in which it was found that 
the long-term cosmic fate depends upon which way the balance tips 
between the forces working to power the cosmic `acceleration', versus 
the opposing factors acting to restrain it. For the former, the only 
real influence working to keep the apparent acceleration going (and 
perhaps ultimately promote it in strength to a real volumetric 
acceleration) was the ever-expanding causal horizon, growing in time, 
out from which an observer can `see' substantial inhomogeneities 
-- i.e., the farthest distance from the observer out to which 
$t_{\mathrm{ret}} > t_\mathrm{Init}$, 
$\Psi (t_{\mathrm{ret}}) \gg 0$ are still true.

For opposition to the acceleration, one source of restraint is 
the FRW expansion of the universe itself, which exerts a natural 
damping effect upon causal backreaction by diluting the density 
of inhomogeneities, simply by pulling them farther away from 
one another (and from any given cosmological observer) over time. 
Also acting to limit a possible long-term acceleration was the 
inevitable saturation presumed for $\Psi (t)$, which -- in our 
original interpretation of this measure of clumping, as described 
in BBI -- would be limited by an upper bound of $\Psi \rightarrow 1$. 

Considering all of these factors, analytical approximations were 
derived for the (pre-RNL) metric perturbation function $I(t)$ as 
$t \rightarrow \infty$; and it was found that $I(t)$ always 
asymptotes to a constant numerical value for $\Psi (t)$ functions 
evolving as a simple power of $t$, $\Psi (t) \propto t^{N}$. 
This asymptotic value of $I(t)$ is larger for smaller $N$, with 
it being equal to unity (implying a complete breakdown of the 
Newtonianly-perturbed metric) for $N = 2/3$ (i.e., 
$\Psi = \Psi _{\mathrm{MD}}$). Thus a fully general-relativistic 
acceleration at late times due to causal backreaction did appear 
to be realistically possible, according to the formalism of BBI 
-- though of course that conclusion was made pending the 
still-undetermined effects of recursive nonlinearities, 
and other complications. 

In this paper, however, we will have to revise our expectations, 
since the incorporation of RNL (recursive nonlinearities) has 
made the prospects of an `eternal', self-powered acceleration 
seem far less likely. We can conclude this despite the fact that 
$I(t)$ is no longer necessarily amenable to a simple analytical 
analysis, even for $t \rightarrow \infty$. One point in favor 
of such a conclusion is that RNL has forced us to change to 
$\Psi (t) \propto t^{N}$ functions with larger exponents $N$ 
in order to re-establish a concordance -- i.e., moving from 
$\Psi _{\mathrm{MD}}$ and $\Psi _{\mathrm{Lin}}$ before, to 
$\Psi _{\mathrm{Sqr}}$ now -- and as our previous results have 
shown, larger exponents in $\Psi (t)$ lead to smaller late-time 
values of $I(t)$ (even if we no longer know its exact asymptotic 
behavior), thus implying a less general-relativistic, less 
`accelerative' long-term evolution. But an even more important 
factor is what we learned about the impact of the ultimate 
`saturation' of the clumping evolution function $\Psi (t)$ 
at some final, fixed numerical value, as was depicted in 
Figure~\ref{FigEarlyVsLateSat}. This result (and other simulation 
runs that we have done along these lines) clearly demonstrate 
that RNL, which acts to stall the expansion of an observer's 
``inhomogeneity horizon'' -- particularly so in cosmologies 
with strong early-time backreaction -- will largely `shut off' 
all apparent acceleration effects not long after the ongoing 
clustering and virialization have ceased, once $\Psi (t)$ has 
settled down to some mostly-constant final value. This behavior 
certainly works against the possibility of an `eternally' 
(or even long-term) `accelerating' universe. 

The one enhancement introduced in this paper which might possibly 
help lead to a long-term acceleration, is the now fundamentally 
unbounded nature of $\Psi (t)$ due to hierarchical clustering 
(assuming that an effective end to clustering is not imposed 
explicitly by early saturation, as we did in fact choose to impose 
for our models in Section~\ref{SubEarlySatClumping}). But while this 
change in interpretation allowing $\Psi (t) > 1$ is indeed based 
upon real physics, is it strong enough to keep the `acceleration' 
going continually, deep into the future? The answer to this question 
is naturally uncertain: on (relatively) small scales, clustering has 
never ceased, as star clusters and small galaxies continue to merge 
(with new virialization) into large galaxies; galaxies continue 
to merge into galactic clusters; and so on. Yet, one would 
realistically expect to find steadily diminishing returns on such 
`small' scales. Any real hope for a continuing acceleration would 
seem to rest upon the future clustering behavior on extraordinarily 
{\it large} scales -- a realm of structure formation with no real 
upper limit, as superclusters eventually manage to self-virialize 
internally, then begin themselves to merge into even larger 
structures, and so on, ad infinitum. Higher and higher scales of 
clustering take exponentially longer to complete than the levels 
below them, though, and it is not clear how large the effective 
`inhomogeneity horizon' for causal backreaction can ever practically 
become -- especially given the fact that the cosmic acceleration 
itself tends to impede the expansion of observational horizons. 
A truly long-term cosmic acceleration (apparent or real) into the 
far future would therefore seem very difficult to accomplish using 
causal backreaction with recursive nonlinearities; though such 
conclusions cannot be considered definitive without more detailed 
cosmological simulations, employing a significantly more 
sophisticated treatment of the effects of vorticity and 
virialization than we have used in these calculations so far, 
given our toy models for $\Psi (t)$, and therefore $I(t)$. 

There are still two remaining wildcards, however, which were 
discussed in BBI and must be mentioned again here. The first one 
is the possible advent of truly general-relativistic perturbations, 
causing the breakdown of our Newtonianly-perturbed metric 
approximation (Equation~\ref{EqnAngAvgBHMatDomWeakFlat}), and the 
failure of our treatment of the individual perturbations due to 
the multitude of locally-clumped masses as being linearly summable. 
The possibility of such a complication still exists, as is made 
obvious (for example) by the very large $I_{0}$ values that we 
typically find for early-saturation runs with large 
$z_\mathrm{Sat}$ settings (e.g., the $z_\mathrm{Sat} = 1$ case 
from Table~\ref{TableRNLearlySatRuns}). As noted above 
in Section~\ref{SubEarlySatClumping}, the fact that 
the propagation of inhomogeneity information obeys the 
proportionality $d r / dt \propto \sqrt{1 - I^{\mathrm{RNL}}(t)}$, 
means that the flow of such information will be choked off whenever 
$I^{\mathrm{RNL}}$ approaches unity, thus freezing the expansion 
of the relevant inhomogeneity horizons, thereby locking 
$I^{\mathrm{RNL}}(t)$ nearly in place at whatever actual value that 
it has managed to grow to by that time. (This is in fact the biggest 
reason why the $z_\mathrm{Sat} = 3$ model from 
Figure~\ref{FigEarlyVsLateSat} experienced such an abrupt shut-off of 
its apparent acceleration so quickly after the growth of its clumping 
evolution function $\Psi (t)$ had ceased.) 
What we cannot know from our formalism, however, is how to physically 
interpret the effects of a metric perturbation function 
$I^{\mathrm{RNL}}(t)$ that is hovering around a nearly fixed value, 
but where that value is always slowly asymptoting towards unity. Does 
it look almost exactly like a decelerating SCDM universe, since a 
constant $I^{\mathrm{RNL}}(t)$ can simply be transformed away through 
coordinate redefinitions? Or would it instead be a universe perpetually 
appearing to be right on the verge of undergoing a runaway acceleration, 
just never quite doing it, basically `riding the edge' forever? (Perhaps 
succeeding, at least, at producing a hesitating, stop-and-go accelerating 
behavior.) Or on the other hand, does the proximity of $I^{\mathrm{RNL}}(t)$ 
to unity manage to override all other considerations, and lead to an actual, 
volumetric, runaway acceleration? The real answers to these questions 
cannot be determined without at least some nonlinear gravitational terms 
being added into the formalism, if not actually requiring a fully 
general-relativistic treatment. But while the quantitative calculation 
of such effects to determine their physical implications is beyond the 
scope of the causal backreaction formalism that has been presented here 
(or in BBI), it is clear that the sum of innumerably many Newtonian-level 
perturbations is indeed capable of driving the total cosmological metric 
perturbation right up to the breaking point, where a `real' cosmic 
acceleration (by any definition) may very possibly take over, 
and for an indeterminate period of time. 

Last of all considerations, though, is the one that can least 
safely be neglected: and that is the eventual breakdown of our 
smoothly-inhomogeneous approximation, as increasing clumpiness 
leads to fundamentally non-negligible anisotropies across much 
of the observable universe itself 
\citep[e.g., the ``Dark Flow'' of][]{KashDarkFlowPersists}; a scenario 
which in BBI was termed the ``Big Mess''. This final, virtually certain 
breakdown of the Cosmological Principle is the ultimate game-changer, 
and questions about the possibility of a `permanent' cosmic 
acceleration due to causal backreaction then become just as 
hard to define as they are to answer, as the fundamental FRW basis 
of cosmological analysis finally breaks down entirely.

\section{\label{SecSummConclude}SUMMARY AND CONCLUSIONS}

In this paper, we have revisited the causal backreaction paradigm 
introduced in \citet{BochnerAccelPaperI}, for which the apparent cosmic 
acceleration is generated not by any form of Dark Energy, but by the 
causal flow of information coming in towards a typical cosmological 
observer from a multitude of Newtonian-strength perturbations, each one 
due to a locally clumped, virializing system. Self-stabilized by vorticity 
and/or velocity dispersion, such perturbations are capable of generating 
positive volume expansion despite their individually-Newtonian natures. 
Noting that previous `no-go' arguments against Newtonian-level backreaction 
are based upon non-causal backreaction frameworks, we see that the sum total 
of these small but innumerable perturbations adds up to an overall effect 
that is strong enough to explain the apparent acceleration as detected by 
Type Ia supernovae, as well as permitting the formulation of an alternative 
cosmic concordance for a matter-only, spatially-flat universe. 

Our purpose here has been to develop and test a second-generation version 
of this causal backreaction formalism, filling in one of the most important 
gaps of the original `toy model' by including what we have termed 
``recursive nonlinearities'' -- specifically referring to the process by 
which old metric perturbation information tends to slow down the causal 
propagation of all future inhomogeneity information, therefore reducing 
the effective cosmological range of causal backreaction effects, and 
thus damping the strength of their overall impact upon the cosmic 
evolution and upon important cosmological observations. 

Utilizing the new simulation program introduced here, which now 
incorporates recursive nonlinearities into causal backreaction, we find 
profound differences in the resulting cosmological model calculations. 
For a given magnitude of self-stabilized clustering assumed for 
large-scale structure, denoted by dimensionless model input parameter 
$\Psi _{0}$, the overall power of causal backreaction is now considerably 
weaker, in addition to fading out relatively rapidly after the growth 
of clustering ceases. This is unlike the results of the original model, 
in which causal backreaction effects would continue to grow regardless 
of any late-time saturation of clustering, due to the causally-expanding 
``inhomogeneity horizon'' seen by an observer which continually brings 
more `old' inhomogeneities into view from ever-greater cosmic distances. 

After discussion of some of the possible reasons for which causal 
backreaction may now appear to fall short of its cosmological goals -- 
either due to issues regarding the fundamental mechanism itself, or due 
to our highly simplified treatment of it -- we then considered a very 
straightforward way in which the paradigm may be fully recovered as a 
cosmological replacement for Dark Energy: all that is needed is the 
adoption of $\Psi _{0}$ values greater than unity. Though representing 
an ad-hoc modification of the original formalism, the change makes 
astrophysical sense in a number of ways. Rather than viewing the 
clumping evolution function $\Psi (t)$ as simply representing the 
fraction of cosmic matter in the `clumped' versus `unclumped' state 
at any given time, $\Psi (t)$ can now be recognized (more realistically) 
as representing the total backreaction effect of hierarchical 
structure formation in the universe, where clustering and virialization 
take place simultaneously on a number of different cosmic length scales -- 
from stellar clusters, to individual galaxies, to galaxy clusters, etc. 
Model input parameter $\Psi _{0} \equiv \Psi (t_{0})$ is now interpreted 
as the effective number of `levels' of completed clustering that exists (at 
current time $t_{0}$) in the large-scale structure when one sums over the 
(partial or total) clustering of matter on all relevant cosmic scales. 

Given this enlarged parameter space with $\Psi _{0} > 1$ now permitted, 
we once again find a selection of model cosmologies that succeed (despite 
the damping effects of recursive nonlinearities) at reproducing the 
observed cosmic acceleration, while also re-establishing an alternative 
cosmic concordance by producing output parameters that match the observables 
derived from several of the most important cosmological data sets. 
Furthermore, astrophysical considerations regarding the necessary {\it input} 
parameters for these apparently successful models -- specifically, the 
need to assume a sufficiently early beginning of clustering -- results in 
a preference by the new formalism for $\Psi (t)$ models that reflect the 
late, nonlinear phase of structure formation. This is an improvement over 
the old formalism without recursive nonlinearities -- which had preferred 
models that embody the early phase of clustering, with linearized matter 
fluctuations -- since it is this final, nonlinear stage of clustering during 
which virialization occurs via the generation of vorticity and velocity 
dispersion, and hence represents the more astrophysically reasonable source 
of substantial causal backreaction. 

Noting that the only problem still remaining for this new concordance is 
the somewhat excessively large amount of clustering required to achieve 
it -- that is, $\Psi _{0} \simeq 4$, rather than what we consider to be 
more reasonable values like $\Psi _{0} \sim 2 - 3$ -- we then determined that 
this problem could be successfully fixed (i.e., a good concordance 
generated with $\Psi _{0} < 3$) by introducing ``early saturation'', 
in which the clumping evolution function $\Psi (t)$ reaches its ultimate 
value of $\Psi _{0}$ somewhat in the past ($z \sim 0.25$), and then changes 
little thereafter. This is a highly reasonable adjustment to the formalism, 
since in the real universe ``gastrophysics'' feedback exists which creates 
superheated baryons, sending large amounts of material back into the 
intergalactic medium, thereby slowing down the continued clustering 
of matter at late times; not to mention the likely slowdown of 
clustering due to the feedback effects of the backreaction, itself. 

The only major drawback of this new feature is the addition of an 
extra model input parameter -- the epoch of saturation, $z_\mathrm{Sat}$ 
-- which results in a degeneracy within $(z_\mathrm{Sat},\Psi _{0})$-space, 
providing a range of models that all fit the Type Ia supernova data well, 
yet lead to significant differences for certain output cosmological parameters. 
The greatest variation in the output results due to this degeneracy occurs 
for the observable jerk parameter, $j_{0}^{\mathrm{Obs}}$, hence implying 
a loss of predictability for $j_{0}^{\mathrm{Obs}}$ by our causal 
backreaction formalism. This is a significant loss, given the previous 
findings from \citet{BochnerAccelPaperI} (without recursive nonlinearities) 
which had indicated that $j_{0}^{\mathrm{Obs}} >> 1$ was the most distinctive 
signature of causal backreaction, thus serving as the clearest way for 
distinguishing it from Cosmological Constant $\Lambda$CDM (or from anything 
close to it), since flat $\Lambda$CDM always requires $j_{0} = 1$. 
It thus becomes more difficult to find a falsifiable test of 
the causal backreaction paradigm, a test that is needed to definitively 
distinguish it from Dark Energy in order to eventually rule out one 
cosmological approach in favor of the other. 

Finally, concerning the `ultimate' fate of the universe, we note that 
the incorporation of recursive nonlinearities tends to shut down any 
strong apparent acceleration effects fairly quickly once the ongoing 
clustering (i.e., the continued growth of $\Psi (t)$) finally stops. 
Even more dramatic is the way in which the metric perturbation function, 
$I^{\mathrm{RNL}}(t)$, becomes essentially locked in place when 
approaching too close to unity, making it an even greater obstacle 
in terms of preventing the acceleration (apparent or otherwise) from 
completely taking over the cosmic evolution. This makes the scenario 
of a perpetual, `eternal' acceleration seem even less likely than it 
already did in \citet{BochnerAccelPaperI}; though the now-unbounded 
nature of $\Psi (t)$ could potentially provide some aid in producing 
a long-term acceleration, as long as virialized structure can continue 
to form on ever-larger cosmic scales, without any fundamental 
upper limit to the sizes of coherent structures. Furthermore, 
the question of the ultimate cosmic fate is once again complicated 
by the possible backreaction contributions of gravitationally 
nonlinear terms, and the (unavoidable) eventual breakdown of the 
approximation of the universe as ``smoothly-inhomogeneous'' -- both 
complications representing scenarios which our toy-model formalism 
is not presently designed to account for. 

In summary, we conclude that our causal backreaction formalism 
remains successful at generating an alternative cosmic concordance 
for a matter-only universe, without requiring any form of Dark Energy; 
though the necessary incorporation of recursive nonlinearities 
into these models implies that a significantly stronger amount of such 
backreaction than before is now needed, acting throughout the crucial 
`acceleration epoch' of $z \sim 0.2 - 2$ or so, in order to provide 
a degree of observed acceleration sufficient to match the 
cosmological standard candle observations.

\acknowledgments

I am grateful to William Chan for computational support.


\newpage

\newpage

\newpage

\newpage

\newpage

\newpage

\newpage


\begin{table}
\begin{center}
\caption{Supernova Fit Quality and New-Concordance Success 
for Optimized RNL Runs 
\label{TableRNLt0SatOptimRuns}}
\scalebox{0.97} 
{
\begin{tabular}{cccc}
\tableline
\tableline
$z_\mathrm{Init}$ & 
$\Psi _{0, \mathrm{Opt}}$ \tablenotemark{a} &
$\chi^{2}_{\mathrm{Fit}}$ \tablenotemark{b} & 
Avg. \% Dev. \tablenotemark{c} 
\\
\tableline
\multicolumn{4}{c}{{\it $\Psi _{\mathrm{MD}}$ Clumping Model Runs}}\\
\tableline
1   & 2.8 & 313.0 & 7.5 \\
{\bf 1.5} & {\bf 2.8} & {\bf 314.7} & {\bf 4.0} \\
{\bf 2} & {\bf 2.9} & {\bf 315.1} & {\bf 5.8} \\
3   & 3.1 & 315.5 & 14.1 \\
5   & 3.6 & 317.0 & 30.4 \\
10  & 4.3 & 328.5 & 52.2 \\
15  & 4.1 & 339.5 & 51.5 \\
20  & 4.1 & 345.8 & 52.6 \\
25  & 4.1 & 349.5 & 53.1 \\
\tableline
\multicolumn{4}{c}{{\it $\Psi _{\mathrm{Lin}}$ Clumping Model Runs}}\\
\tableline
1   & 3.2 & 311.0 & 7.8 \\
1.5 & 3.1 & 312.8 & 5.6 \\
{\bf 2} & {\bf 3.2} & {\bf 313.6} & {\bf 2.9} \\
{\bf 3} & {\bf 3.3} & {\bf 314.0} & {\bf 4.8} \\
5   & 3.6 & 314.3 & 11.9 \\
10  & 4.0 & 314.7 & 21.4 \\
15  & 4.3 & 314.9 & 26.7 \\
20  & 4.5 & 315.0 & 30.1 \\
25  & 4.7 & 315.1 & 33.0 \\
\tableline
\multicolumn{4}{c}{{\it $\Psi _{\mathrm{Sqr}}$ Clumping Model Runs}}\\
\tableline
1   & 4.4 & 311.5 & 15.6 \\
1.5 & 4.0 & 310.4 & 12.0 \\
2   & 3.9 & 310.5 & 10.0 \\
3   & 3.9 & 311.0 & 7.6 \\
5   & 4.0 & 311.4 & 5.4 \\
10  & 4.1 & 311.7 & 3.4 \\
15  & 4.1 & 311.8 & 2.9 \\
20  & 4.1 & 311.8 & 2.6 \\
{\bf 25} & {\bf 4.1} & {\bf 311.8} & {\bf 2.5} \\
\tableline
\end{tabular}
}
\tablenotetext{a}{Each $\Psi _{0, \mathrm{Opt}}$ chosen 
to minimize $\chi^{2}_{\mathrm{Fit}}$ for a given 
$z_\mathrm{Init}$ and $\Psi (t)$ function.}
\tablenotetext{b}{$\chi^{2}_{\mathrm{Fit}}$ computed 
versus the SCP Union1 SNIa data set. 
}
\tablenotetext{c}{Average Percent Deviation from the 
``Cosmic Concordance'', defined (as described in the text) 
via cosmological parameters obtained from the 
Union1-best-fit $\Lambda$CDM model 
($\Omega _{\Lambda} = 0.713 = 1 - \Omega _\mathrm{M}$, 
$\chi^{2}_{\mathrm{Fit}} = 311.9$).}
\end{center}
\end{table}

\newpage

\begin{table}
\begin{center}
\caption{Output Cosmological Parameters from our `Best' Runs with Recursive Nonlinearities
\label{TableRNLt0SatBestRuns}}
\resizebox{16.5cm}{!} 
{
\begin{tabular}{ccccccccccccc}
\tableline
\tableline
$z_\mathrm{Init}$ & 
$\Psi _{0}$ &
$\chi^{2}_{\mathrm{Fit}}$ \tablenotemark{a} & 
$P_{\mathrm{Fit}}$ \tablenotemark{b} & 
$I_{0}$ \tablenotemark{c} &
$z^\mathrm{Obs}$ \tablenotemark{d} & 
$H^\mathrm{Obs}_{0}$ \tablenotemark{e} & 
$H^\mathrm{FRW}_{0}$ \tablenotemark{f} & 
$t^\mathrm{Obs}_{0}$ \tablenotemark{g} & 
$\Omega^\mathrm{FRW}_\mathrm{M}$ \tablenotemark{h} &
$w^\mathrm{Obs}_{0}$ & 
$j^\mathrm{Obs}_{0}$ & 
$l^\mathrm{Obs}_{\mathrm{A}}$ 
\\
\tableline
\multicolumn{13}{c}{{\it $\Psi _{\mathrm{MD}}$ Clumping Model Runs}}\\
\tableline
1.5 & 2.8 & 314.7 & 0.324 & 0.53 & 1.16 & 69.54 & 41.96 & 
13.95 & 0.948 & -0.639 & 0.59 & 297.8 \\
2 & 2.9 & 315.1 & 0.319 & 0.62 & 1.17 & 69.58 & 38.62 & 
14.44 & 1.162 & -0.639 & 0.51 & 288.7 \\
\tableline
\multicolumn{13}{c}{{\it $\Psi _{\mathrm{Lin}}$ Clumping Model Runs}}\\
\tableline
2 & 3.2 & 313.6 & 0.340 & 0.53 & 1.16 & 69.75 & 42.10 & 
13.88 & 0.946 & -0.675 & 0.89 & 297.3 \\
3 & 3.3 & 314.0 & 0.334 & 0.62 & 1.16 & 69.68 & 38.72 & 
14.30 & 1.159 & -0.667 & 0.83 & 287.2 \\
\tableline
\multicolumn{13}{c}{{\it $\Psi _{\mathrm{Sqr}}$ Clumping Model Run}}\\
\tableline
25 & 4.1 & 311.8 & 0.367 & 0.53 & 1.14 & 70.07 & 42.32 & 
13.64 & 0.943 & -0.751 & 1.73 & 294.5 \\
\tableline
\multicolumn{13}{c}{{\it Comparison Values from 
Best-Fit \tablenotemark{i} 
flat $\Lambda{\mathrm{CDM}}$ Model 
$(\Omega _{\Lambda} = 0.713 = 1 - \Omega _\mathrm{M})$}} 
\\
\tableline
\nodata & \nodata & 311.9 & 0.380 & \nodata & 1.0 & 69.96 & 69.96 & 
13.64 & 0.287 & -0.713 & 1.0 & 285.4 \\
\tableline
\multicolumn{13}{c}{{\it Comparison Values from 
Best-Fit \tablenotemark{j} flat SCDM Model 
$(\Omega _{\Lambda} = 0$, $\Omega _\mathrm{M} = 1)$}}\\
\tableline
\nodata & \nodata & 608.2 & 3.4E-22 & \nodata & 1.0 & 61.35 & 61.35 & 
10.62 & 1.0 & 0.0 & 1.0 & 287.3 \\
\tableline
\end{tabular}
}
\tablenotetext{a}{$\chi^{2}_{\mathrm{Fit}}$ computed versus the 
SCP Union1 SNIa data set.} 
\tablenotetext{b}{Each likelihood probability $P_{\mathrm{Fit}}$ 
is derived from the corresponding 
$\chi^{2}_{\mathrm{Fit}}$ using the $\chi^{2}_{N_{\mathrm{DoF}}}$ 
distribution with $N_{\mathrm{DoF}}$ degrees of freedom, where 
$N_{\mathrm{DoF}} = 304$ for our $\Psi _{\mathrm{MD}}$, 
$\Psi _{\mathrm{Lin}}$, and $\Psi _{\mathrm{Sqr}}$ clumping models, 
$N_{\mathrm{DoF}} = 305$ for the flat $\Lambda{\mathrm{CDM}}$ model, 
and $N_{\mathrm{DoF}} = 306$ for flat SCDM.}
\tablenotetext{c}{The integrated (Newtonian) gravitational 
perturbation potential at $t_{0}$, as modified by RNL, computed via 
Equations~\ref{EqnItotRNLintegration1}-\ref{EqnRNLdiscreteGridLoopB}.}
\tablenotetext{d}{Each $z^\mathrm{Obs}$ quoted here corresponds 
to $z^\mathrm{FRW} \equiv 1$.}
\tablenotetext{e}{The $H^\mathrm{Obs}_{0}$ value 
(given here in $\mathrm{km} ~ \mathrm{s}^{-1} \mathrm{Mpc}^{-1}$) 
for each run is found by minimizing its $\chi^{2}_{\mathrm{Fit}}$ 
with respect to the SCP Union1 SNIa data set.}
\tablenotetext{f}{Each $H^\mathrm{FRW}_{0}$ is computed relative 
to the corresponding optimized $H^\mathrm{Obs}_{0}$ value 
for that run.}
\tablenotetext{g}{All $t^\mathrm{Obs}_{0}$ values are listed here 
in GYr, and computed assuming {\it no radiation} 
(i.e., $\Omega _{R} \equiv 0$).}
\tablenotetext{h}{All $\Omega^\mathrm{FRW}_\mathrm{M}$ values given 
here for the $\Psi _{\mathrm{MD}}$, $\Psi _{\mathrm{Lin}}$, and 
$\Psi _{\mathrm{Sqr}}$ models are normalized to 
$\Omega^\mathrm{Obs}_\mathrm{M} \equiv 0.27$.}
\tablenotetext{i}{``Best-Fit" for the flat 
$\Lambda{\mathrm{CDM}}$ Model refers here to an optimization 
over $\Omega _{\Lambda}$ and $H^\mathrm{Obs}_{0}$.}
\tablenotetext{j}{``Best-Fit" for the flat 
SCDM Model refers here to an optimization 
over $H^\mathrm{Obs}_{0}$.}
\end{center}
\end{table}

\newpage

\begin{table}
\begin{center}
\caption{Output Cosmological Parameters from our RNL Runs with `Early Saturation' 
\label{TableRNLearlySatRuns}}
\resizebox{16.5cm}{!} 
{
\begin{tabular}{ccccccccccccc}
\tableline
\tableline
$z_\mathrm{Sat}$ & 
$\Psi _{0, \mathrm{Opt}}$ \tablenotemark{a} & 
$\chi^{2}_{\mathrm{Fit}}$ & 
$P_{\mathrm{Fit}}$ \tablenotemark{b} & 
$I_{0}$ &
$z^\mathrm{Obs}$ & 
$H^\mathrm{Obs}_{0}$ & 
$H^\mathrm{FRW}_{0}$ & 
$t^\mathrm{Obs}_{0}$ & 
$\Omega^\mathrm{FRW}_\mathrm{M}$ &
$w^\mathrm{Obs}_{0}$ & 
$j^\mathrm{Obs}_{0}$ & 
$l^\mathrm{Obs}_{\mathrm{A}}$ 
\\
\tableline
\multicolumn{13}{c}{{\it $\Psi _{\mathrm{Sat}}$ Clumping Model Runs, 
$z_\mathrm{Init} = 25$}}\\
\tableline
0 & 4.1 & 311.8 & 0.351 & 0.53 & 1.14 & 70.07 & 42.32 & 
13.64 & 0.943 & -0.751 & 1.73 & 294.5 \\
0.25 & 2.6 & 313.5 & 0.326 & 0.58 & 1.15 & 69.60 & 40.24 & 
14.00 & 1.054 & -0.620 & 0.15 & 289.7 \\
0.5 & 2.3 & 316.6 & 0.284 & 0.68 & 1.15 & 69.40 & 36.32 & 
14.65 & 1.338 & -0.585 & -0.14 & 279.8 \\
1 & 2.2 & 320.2 & 0.238 & 0.80 & 1.14 & 68.77 & 29.54 & 
15.75 & 2.086 & -0.488 & -0.94 & 259.9 \\
\tableline
\multicolumn{13}{c}{{\it Comparison Values from 
Best-Fit flat $\Lambda{\mathrm{CDM}}$ Model 
$(\Omega _{\Lambda} = 0.713 = 1 - \Omega _\mathrm{M})$}} 
\\
\tableline
\nodata & \nodata & 311.9 & 0.380 & \nodata & 1.0 & 69.96 & 69.96 & 
13.64 & 0.287 & -0.713 & 1.0 & 285.4 \\
\tableline
\end{tabular}
}
\tablenotetext{a}{Each $\Psi _{0, \mathrm{Opt}}$ is chosen to minimize 
$\chi^{2}_{\mathrm{Fit}}$ for a given $z_\mathrm{Sat}$, with 
$z_\mathrm{Init} = 25$ and $\Psi (t) \equiv \Psi _{\mathrm{Sat}} (t)$. 
(Optimizations are performed with respect to the SCP Union1 
SNIa data set.)}
\tablenotetext{b}{Each likelihood probability $P_{\mathrm{Fit}}$ 
is derived from the corresponding $\chi^{2}_{\mathrm{Fit}}$ using 
the $\chi^{2}_{N_{\mathrm{DoF}}}$ distribution with $N_{\mathrm{DoF}}$ 
degrees of freedom, where the inclusion of optimizable parameter 
$z_\mathrm{Sat}$ now gives $N_{\mathrm{DoF}} = 303$ for our 
$\Psi _{\mathrm{Sat}}$ clumping models (even for $z_\mathrm{Sat} = 0$), 
while $N_{\mathrm{DoF}} = 305$ is still used here for the flat 
$\Lambda{\mathrm{CDM}}$ model.}
\end{center}
\end{table}

\end{document}